\RequirePackage{fix-cm}
\documentclass[twocolumn,epjc3,nopacs]{svjour3}

\RequirePackage{graphicx} 
\usepackage{placeins}
\usepackage{bm}
\usepackage{epsfig}
\usepackage[font={small},labelsep=quad]{caption}
\usepackage{subfig} 
\usepackage[export]{adjustbox}
\usepackage{multirow}
\usepackage{cite}
\usepackage[colorlinks=true,citecolor=blue,filecolor=blue,linkcolor=blue,urlcolor=blue,pdftex]{hyperref}
\usepackage{verbatim}
\usepackage{mathtools}
\usepackage{mhchem}
\usepackage{siunitx}
\usepackage{supertabular,booktabs}
\usepackage{lineno}
\usepackage[usenames,dvipsnames]{xcolor}
\usepackage[normalem]{ulem}

\definecolor{forestgreen}{rgb}{0.13, 0.55, 0.13}

\journalname{Eur. Phys. J. C}

\begin{document}
\title{The first dual-phase xenon TPC equipped with silicon photomultipliers and characterisation with \ce{^{37}Ar}}

\author{L.~Baudis
\and Y.~Biondi
\and M.~Galloway
\and F.~Girard
\and S.~Hochrein
\and S.~Reichard
\and P.~Sanchez-Lucas\thanksref{e2}
\and K.~Thieme\thanksref{e1}
\and J.~Wulf
}

\institute{\normalsize{Department of Physics, University of Zurich, Winterthurerstrasse 190, CH-8057, Z\"{u}rich, Switzerland}}

\thankstext{e2}{e-mail: patricia.sanchez@physik.uzh.ch}
\thankstext{e1}{e-mail: kevin.thieme@physik.uzh.ch}

\date{}

\maketitle

\sloppy

\begin{abstract} 
For the first time, a small dual-phase (liquid/gas) xenon time projection chamber was equipped with a top array of silicon photomultipliers for light and charge readout. Here we describe the instrument in detail, as well as the data processing and the event position reconstruction algorithms. We obtain a spatial resolution of $\sim \SI{1.5}{mm}$ in the horizontal plane. To characterise the detector performance, we show calibration data with internal \ce{^{83\text{m}}Kr} and \ce{^{37}Ar} sources, and we detail the production of the latter as well as its introduction into the system. We finally compare the observed light and charge yields down to electronic recoil energies of $\SI{2.82}{keV}$ to predictions based on NEST~v2.0.
\end{abstract}

\section{Introduction}
\label{sec:intro}
Detectors that use liquid xenon are widely employed in the field of astroparticle physics. In particular, dual-phase (liquid/gas) time projection chambers (TPCs) are used for direct dark matter detection, searches for the neutrinoless double beta decay of \ce{^{136}Xe}, as well as other rare event searches~\cite{Chepel:2012sj,Aprile:1900zz}.

A defining requirement to observe a rare interaction is an ultra-low background rate, to maximise the signal-to-background ratio and thus the sensitivity to a given search channel. This is achieved via the reduction of the radioactivity levels of detector materials, the definition of central detector regions via fiducialisation to benefit from the self-shielding power of liquid xenon, and the distinction between single and multiple interactions. The latter rely on precise position reconstruction capabilities of the interaction sites.

In a dual-phase TPC, particle interactions are detected by observing prompt scintillation light (S1) and a delayed (S2) signal caused by ionisation electrons, drifted and extracted into the gaseous phase where they produce electroluminescence. Both S1 and S2 signals are observed by photosensors. The depth of an interaction is given by the drift time between S1 and S2, while the $(x,y)$-position is derived from the light distribution in the photosensor plane placed in the gas phase above the liquid.

Many past dark matter searches based on liquid xenon TPCs~\cite{Akerib:2012ys,Aprile:2017aty,Cui:2017nnn} and experiments under construction~\cite{Mount:2017qzi,Zhang:2018xdp} employ VUV-sensitive photomultiplier tubes (PMTs)  which have been optimised for low levels of radioactivity and are up to $\SI{3}{inches}$ in diameter~\cite{Akerib:2012ys,aprile:2015lha,Baudis:2013xva,Li:2015qhq}. However, they are among the dominant sources of electronic (ER) and nuclear recoil (NR) backgrounds, and they limit the $(x,y)$-position resolution, which is correlated to their size. Next-generation detectors such as DARWIN~\cite{Aalbers:2016jon} and nEXO~\cite{Kharusi:2018eqi,Jamil:2018tkx} aim to reduce the backgrounds further, and consider the use of silicon photomultipliers (SiPMs) for one or both photosensor planes.

Motivated by these future experiments, and by our previous studies of VUV-sensitive SiPM arrays in single-phase detectors~\cite{Baudis:2018pdv}, we have built and characterised the first dual-phase xenon TPC equipped with a top photosensor array containing 16~SiPM channels read out individually. The new detector is based on our previously described Xurich~II TPC~\cite{Baudis:2017xov}, with several modifications as detailed below. The calibration of the detector, i.e.~the determination of its charge and light response parameters was performed with two sources, \ce{^{83\text{m}}Kr} and \ce{^{37}Ar}, which allow us to investigate ER interactions at low energies. 

This article is structured as follows: We describe the detector in Section~\ref{sec:detector} and the \ce{^{37}Ar} calibration source in Section~\ref{sec:argon}. We detail the data taking and processing tools in Section~\ref{sec:methods}, and show the analysis of data acquired with both internal calibration sources in Section~\ref{sec:analysis}. We present and discuss the results in Section~\ref{sec:results}, and summarise the main findings, addressing some challenges facing future detectors in Section~\ref{sec:summary}.

\section{The Xurich II detector with SiPM readout}
\label{sec:detector}
The current TPC is an upgrade of the Xurich~II detector operated at the University of Zurich (UZH) and described in~\cite{Baudis:2017xov}. The upgrade focuses on two aspects: the replacement of the top PMT with an array of SiPMs and their read-out electronics, and the design of a stand-alone setup to introduce the gaseous  \ce{^{37}Ar} calibration source into the xenon recirculation loop.

\begin{figure}[htp!]
\centering
\includegraphics*[width=0.5\textwidth]{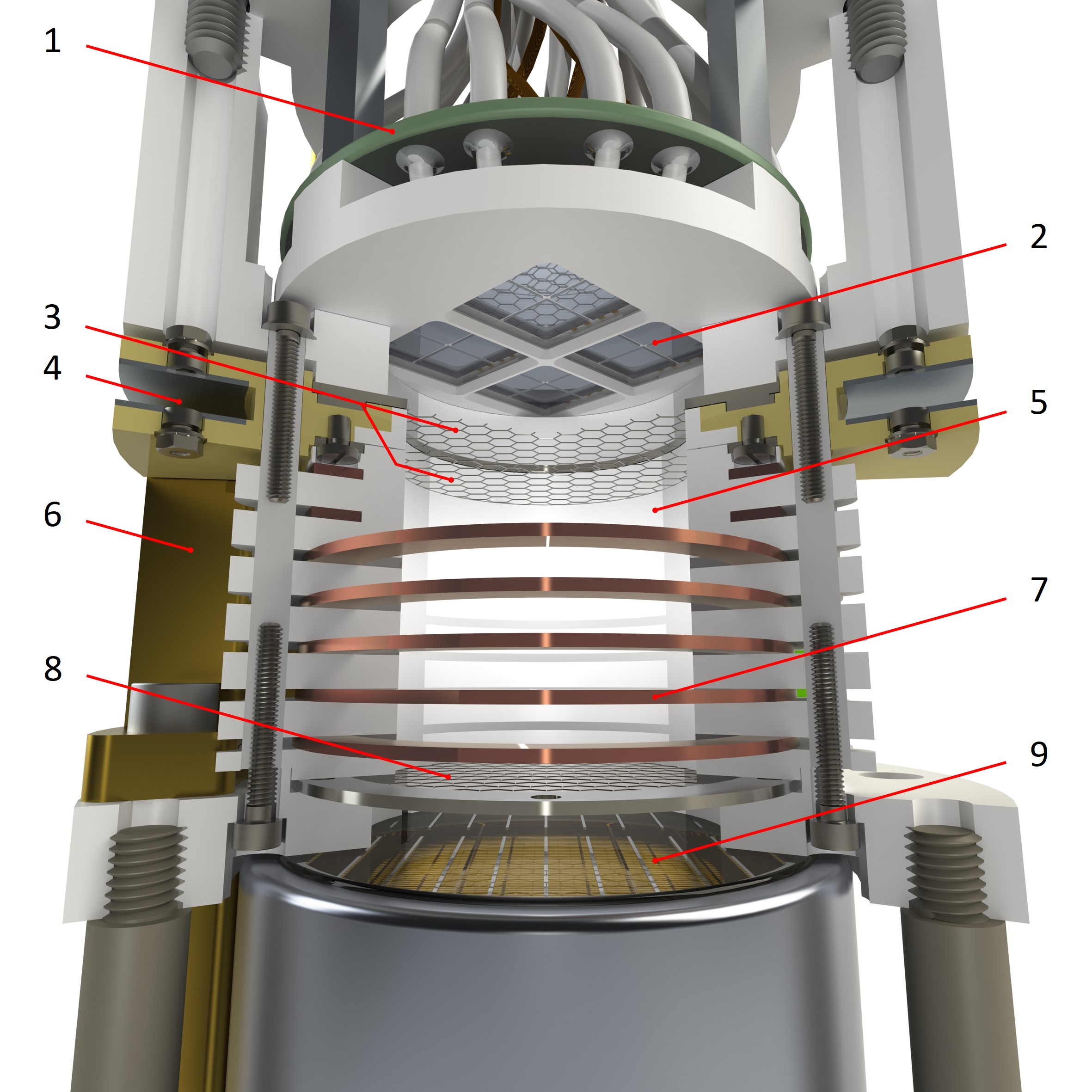}
\caption{Rendering of the upgraded Xurich~II TPC. The active liquid xenon (LXe) is contained within a PTFE cylinder and surrounded by copper field-shaping rings. A \mbox{2-inch} PMT is placed in the liquid, and an array of 16~SiPMs is placed in the gaseous phase at the top. For the sake of visualisation, two field-shaping rings have been cut. Legend: \mbox{1 -- PCB} with $\times 10$~preamplifier, \mbox{2 -- SiPM} array, \mbox{3 -- Anode} and gate mesh, \mbox{4 -- Level} meters, \mbox{5 -- PTFE} reflector wall, \mbox{6 --  Weir} for liquid level control, \mbox{7 -- Copper} field-shaping rings, \mbox{8 -- Cathode}, \mbox{9 -- PMT}.}
\label{fig:TPC}
\end{figure}

The detector is placed inside a vacuum-insulated stainless steel vessel which is coupled to a liquid nitrogen bath via a copper cold finger. The TPC drift region is defined by a $\SI{31}{mm} \times \SI{31}{mm}$ (diameter~$\times$~height) polytetrafluoroethylene (PTFE) cylindrical shell with a cathode mesh at the bottom and a gate mesh at the top, shown in Figure~\ref{fig:TPC}. The anode is placed $\SI{4}{mm}$ above the gate. The liquid surface was kept in the middle of the two top electrodes, at $\sim \SI{2}{mm}$ above the gate for the presented data, optimised for S2 amplification. Seven copper field-shaping rings, separated by PTFE spacers and connected via a resistor chain, ensure a uniform drift field, which is maintained between the negatively biased cathode and the gate at ground potential. By varying the cathode voltage, we can acquire data at different drift fields. For the \ce{^{83\text{m}}Kr} source, we acquired data in the range $\SIrange{194}{968}{V/cm}$ while we probed the interval $\SIrange{80}{968}{V/cm}$ for \ce{^{37}Ar}. The electron extraction field is created between the gate and the positively biased anode and was kept constant at $\SI{10}{kV/cm}$ by setting the anode to a nominal voltage of $+\SI{4}{kV}$.

The xenon is continuously purified of electronegative impurities by circulating the gas phase through a hot metal getter, an integral part of the gas handling system. The gas system features a source chamber which allows for the introduction of \ce{^{83\text{m}}Kr} from the decay of \ce{^{83}Rb} with $T_{1/2}=\SI{86.2 (1)}{\day}$~\cite{Baudis:2017xov,McCutchan:2015}. With an additional gas mixing setup, described in Section~\ref{sec:argon}, that can be connected in parallel to the \ce{^{83}Rb} source chamber, we also deploy a gaseous \ce{^{37}Ar} source.

The S1 and S2 signals are detected by a \mbox{2-inch} diameter PMT (R6041-06 MOD, Hamamatsu Photonics), placed in the liquid phase at the bottom of the TPC, and sixteen $\SI[product-units = single]{6 x 6}{mm^2}$~VUV4 SiPMs ($2 \times 2$ array of S13371, Hamamatsu Photonics) placed in the gas phase. To install the SiPMs in the TPC, a PTFE holder which exactly replaced the area occupied by the former top PMT was designed and built. The SiPMs were arranged in the holder to maximise the light collection efficiency (LCE) as studied in MC simulations~\cite{WulfThesis}. The distance from the surface of the SiPMs to the anode is $\SI{5}{mm}$, optimised to obtain good position resolution.

The new configuration of the TPC requires 17 readout channels, 15 more than in the previous design. To handle this number of cables in the small detector, we installed a new CF-40 feedthrough for 20 double-shielded coaxial signal cables (RG~196) for the photosensors and 7 Kapton\textsuperscript{\textregistered} insulated wires for voltage biasing. 
To independently read out the 16~SiPM channels, a printed circuit board (PCB) was designed. Each SiPM, on its cathode side, is connected in series with a $\SI{10}{k\Omega}$ resistor to the bias voltage to limit the current and avoid possible damage to the device. On its anode side, each SiPM is grounded with a $\SI{50}{\Omega}$ resistor and coupled to a custom-made $\times 10$~voltage amplifier circuit~\cite{WulfThesis}. The main component of the circuit is the ultra-low noise, non-inverting voltage feedback operational amplifier OPA847 from Texas Instruments. This amplifier has a nominal gain of 20 when connected to an open circuit at the output and is able to reduce the electronic pick-up noise to a negligible level. It operates from direct current to $\SI{250}{MHz}$ and has a $\SI{50}{\Omega}$ output impedance that matches the input impedance of the ADC. These two resistances form a voltage divider at the amplifier output. Because of this loading effect, the effective final signal amplification with a $\SI{50}{\Omega}$ load from the ADC is tenfold. The heat dissipation of the entire amplifier board is $\sim \SI{3}{W}$, whereas it was negligible for the former 2-inch PMT base. This additional heat load can safely be handled by the cooling system of the detector.
Located in the warmer gas phase at $\sim \SI{190}{K}$, the SiPMs have shown an error-weighted mean dark count rate of $(\SI{8.05}{}\pm \SI{0.03}{})\si{\, Hz/mm^2}$ at a bias voltage of $\SI{51.5}{V}$, which is consistent with the temperature behaviour we reported~\cite{Baudis:2018pdv}. The dark count rate is the uncorrelated noise rate above a threshold of $\SI{0.5}{photoelectrons}$~(PE) in scintillation-free data acquired with nitrogen gas in the TPC under very similar thermodynamic conditions to those observed when using xenon.

\section{The \ce{^{37}Ar} calibration source}
\label{sec:argon}
The \ce{^{37}Ar} isotope decays with $T_{1/2}=\SI{35.01 (2)}{\day}$ via electron capture into stable \ce{^{37}Cl}, with a Q-value of $\SI{813.9 (2)}{keV}$~\cite{Cameron:2012}. Depending on the inner atomic shell from which the electron is captured, $\SI{2.82}{keV}$, $\SI{0.27}{keV}$ and $\SI{0.0175}{keV}$ X-rays and Auger electrons are released \mbox{(K-,} \mbox{L-}, and \mbox{M-shell}, respectively, see Table~\ref{tab:Ar} for the corresponding branching ratios).
\begin{table}[h]
  \centering
  \begin{tabular}{ccc}
    \toprule
    Decay mode & Energy release $[\si{keV}]$ & Branching ratio \\
    \midrule
    K-capture & $2.8224$ & $90.2\%$ \\
    L-capture & $0.2702$ & $8.9\%$ \\
    M-capture & $0.0175$ & $0.9\%$ \\   
    \bottomrule
  \end{tabular}
  \caption{Energy release and branching ratios of decay modes of \ce{^{37} Ar}~\cite{Barsanov:2007}.}
  \label{tab:Ar}
\end{table}
These offer mono-energetic lines with uniformly distributed events in the TPC that can be used for calibrations down to low ER energies~\cite{Boulton:2017,LUX:2019,Akimov:2014}.

\subsection{Source production}
The \ce{^{37}Ar} source was produced by thermal neutron activation, where the process \ce{^{36}Ar} + n $ \to $ \ce{^{37}Ar} + $\gamma$ with a capture cross section of $\sim \SI{5}{b}$ was exploited. We used natural argon gas with a $0.334 \,\%$ \ce{^{36}Ar} abundance. The activation took place between the 4th and 7th~December~2018 at the Swiss Spallation Source (SINQ) at Paul Scherrer Institute (PSI), Villigen (Switzerland), providing a thermal neutron flux of $\sim \SI{e13}{{cm^{-2}s^{-1}}}$. A total of four fused quartz ampules, made at the glassblowing facility at UZH and filled with natural argon~6.0 (purity $\geq 99.9999 \, \%$) at a pressure of $\SI{0.8}{bar}$ or $\SI{0.9}{bar}$, corresponding to $\SI{2.3}{mg}$ and $\SI{2.7}{mg}$ respectively, were each activated for an irradiation time of 3.75 hours. Analytically, we expect an initial \ce{^{37}Ar} activity between $\SIrange{19}{22}{kBq}$ per ampule. The use of natural argon gives rise to the activation of two other relevant isotopes: \ce{^{39}Ar} from \ce{^{38}Ar} and \ce{^{41}Ar} from \ce{^{40}Ar}. Of less relevance are \ce{^{43}Ar} from \ce{^{42}Ar}, of which only traces are present in natural argon, and \ce{^{42}Ar} itself from secondary activation of \ce{^{41}Ar}. 

As the natural abundance of \ce{^{38}Ar} is more than five times lower and its cross section for thermal neutrons is more than six times lower, we expect to have produced \ce{^{39}Ar} with an activity lower than $\SI{0.23}{Bq}$ per ampule. It decays with a half-life of $\SI{268}{y}$ into its stable daughter \ce{^{39}K} solely via beta decay with an endpoint of $\SI{565}{keV}$~\cite{Chen:2018}. 

The high abundance of $99.6 \,\%$ of \ce{^{40}Ar} yields a high activity of \ce{^{41}Ar} shortly after the irradiation. This required that the activation happened in two steps with a cool down time in between. \ce{^{41}Ar} decays into stable \ce{^{41}K} with a half-life of $\SI{109.6}{min}$~\cite{Nesaraja:2016}. Hence, its activity decreased below $\SI{1}{Bq}$ after 2 days and was negligible after $\sim 160$ days when the source was introduced into the detector. Similarly, if \ce{^{43}Ar} ($T_{1/2}=\SI{5.4}{min}$) had been produced, it would have been negligible after a short time as the subsequent decay of its daughter isotope \ce{^{43}K} into stable \ce{^{43}Ca} takes place with a half-life of $\SI{22.3}{h}$~\cite{Singh:2015}. The secondary activation of \ce{^{41}Ar} to \ce{^{42}Ar} ($T_{1/2}=\SI{32.9}{y}$), which decays via \ce{^{42}K} into stable \ce{^{42}Ca}~\cite{Singh2011}, is negligible as well since the initial activity can be calculated to be $\sim \SI{50}{\micro Bq}$.

\subsection{Source introduction setup}
\begin{figure}[h]
\centering
\includegraphics*[width=0.5\textwidth]{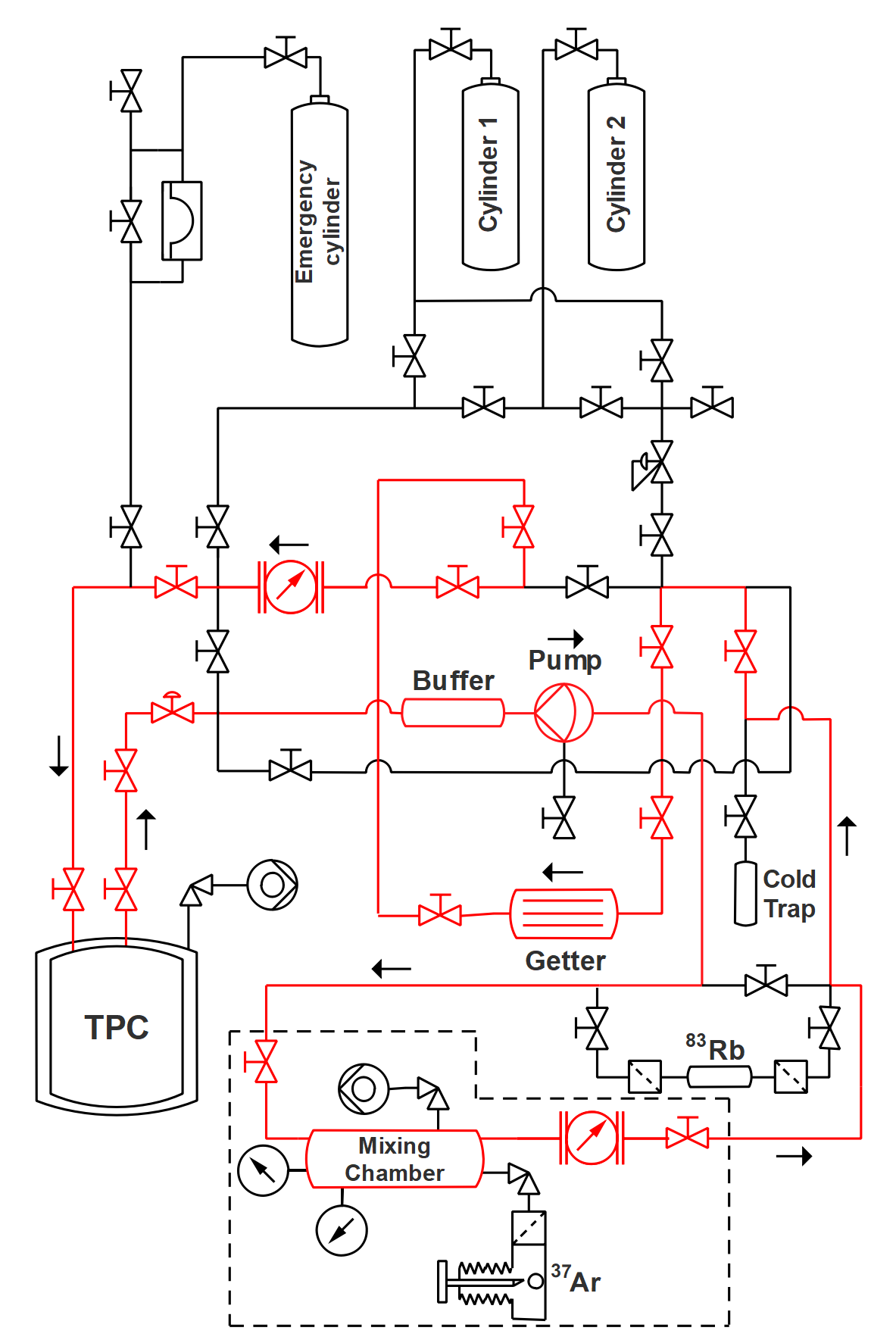}
\caption{Piping and instrumentation diagram of the gas system and the added \ce{^{37}Ar} setup (framed with a dashed line). The gas recirculation path to introduce the source into the xenon flow is indicated in red.}
\label{fig:Gassystem_PID}
\end{figure}

To introduce the \ce{^{37}Ar} source into the TPC, the quartz ampules must be placed and broken inside the gas system. We designed and built a dedicated system that contains a mixing chamber and an ampule breaking mechanism similar to the one described in~\cite{Hils:2014} and deployed in XENON1T at the end of its lifetime~\cite{AlfonsiTalk:2019}. We refer to Figure~\ref{fig:Gassystem_PID} for a schematic view.

The fragile ampule is firmly held in place by a custom-made holder welded onto a CF-40 blank mounted on one of the sides of a CF-Tee. By actuating a guillotine through a vacuum bellow, the neck of the ampule can be cut off in vacuum. Opening a valve to the evacuated mixing volume transfers $\sim 94 \, \%$ of the argon through a filter. After isolating the ampule chamber again, xenon from a gas bottle can be added until the TPC working pressure of $\sim \SI{2}{bar}$ is reached. Bypassing the recirculation flow through the mixing chamber and integrating the flow, the introduced \ce{^{37}Ar} activity can be estimated. The argon gas of multiple ampules can be injected by stopping the recirculation, isolating the TPC and recuperating the xenon/argon mixture remaining in the gas system into a cold trap cooled by liquid nitrogen. The gas system can be pumped after inserting a new ampule from a port on the mixing chamber (cf.~Figure~\ref{fig:Gassystem_PID}). The use of the cold trap ensures that the amount of xenon and, hence, the pressure and liquid level remain unchanged during ampule changes. For the data described in the subsequent sections, three ampules with an estimated total activity of $\SI{2.7}{kBq}$ were broken mid-May~2019 to release the activated argon into the xenon. 

\section{Data toolkit}
\label{sec:methods}
\subsection{Data acquisition}
 The signal sizes are energy dependent, thus deploying different calibration sources requires an adjustment of the settings in the data acquisition (DAQ). The S1~signals of the \ce{^{37}Ar} lines, for instance, are much smaller than those of the \ce{^{83\text{m}}Kr} lines and can even be missed by the trigger or absent in the waveform. At the high end in signal size, the PMT channel could, if not limited in voltage, saturate the electronics with S2~signals from \ce{^{83\text{m}}Kr}. Below, we discuss the DAQ settings and the deployed hardware in detail.

The raw data is digitised by three \mbox{8-}channel CAEN~V1724 modules connected in daisy chain, each with a $\SI{100}{MHz}$ sampling rate ($\SI{10}{ns}$ samples) and \mbox{14-}bit resolution at $\SI{2.25}{V}$ dynamic range. The pre-amplified SiPM signals are fed directly to the Flash ADCs. The PMT signal is attenuated by a factor of 10 for the \ce{^{83\text{m}}Kr} data, while not attenuated for \ce{^{37}Ar}, then sent to the digitisers. After passing through a CAEN~625 fan-in/fan-out module the trigger is generated by a CAEN~N840 leading edge discriminator based on the PMT signal. For \ce{^{83\text{m}}Kr}, we require the attenuated PMT signal to exceed $\SI{2}{mV}$, which corresponds to a peak with integrated charge of $> \SIrange{15}{21}{PE}$. For \ce{^{37}Ar}, we trigger on $\SI{7}{mV}$ of the raw PMT signal corresponding to $> \SIrange{4}{5}{PE}$. Only for the data at $\SI{80}{V/cm}$ drift field we use $\SI{6}{mV}$.
The event window for the \ce{^{83\text{m}}Kr} data was chosen to be $\SI{40}{\micro s}$ with $\SI{30}{\micro s}$ post-trigger time, which is sufficiently large to contain the S2 even for the lowest field of $\SI{194}{V/cm}$ when triggered on the S1. Triggers on \ce{^{83\text{m}}Kr} S2 signals are expected to be rare and are removed in the analysis phase to avoid a bias of events towards the upper half of the TPC. 
For \ce{^{37}Ar}, the event window was chosen to be $\SI{60}{\micro s}$ with the trigger placed in the middle at $\SI{30}{\micro s}$ to contain the entire waveform, regardless of whether the trigger was issued by an S1 or S2. The maximum drift time at the lowest applied field of $\SI{80}{V/cm}$ inside the TPC volume is $\sim \SI{22}{\micro s}$.

Depending on the user settings, entire waveforms can be written to disk or, to save memory, only the samples exceeding a certain threshold (henceforth referred to as ``good'' as distinguished from ``skipped'') by means of Zero Length Encoding (ZLE), as shown in Figure~\ref{fig:Wf_Proc}. The first half of the \ce{^{83\text{m}}Kr} data was taken without ZLE while the rest, including the \ce{^{37}Ar} data, was acquired with ZLE. The ZLE threshold for \ce{^{83\text{m}}Kr}, which only comes into play if a trigger is issued, was chosen to be 15 bins ($\sim \SI{2.06}{mV}$ at ADC input) for the PMT channel and 10 bins ($\sim \SI{1.37}{mV}$ at ADC input) for the SiPM channels. To catch the S1s of \ce{^{37}Ar} at the single PE level, we have chosen 20 bins ($\sim \SI{2.75}{mV}$ at ADC input) for the PMT channel and 6 bins ($\sim \SI{0.82}{mV}$ at ADC input) for the SiPM channels as ZLE thresholds. These values were determined empirically and, in the case of \ce{^{37}Ar}, set just above the baseline. In addition to the over-threshold samples, a certain number of backward- and forward-samples can be specified to include the rise and the tail of the peak. To keep timing information, the counts of good samples and skipped samples are recorded as so-called \emph{control words}~\cite{CAEN:2017}. We recorded 150~(200) backward-samples and 150~(100) forward-samples for \ce{^{83\text{m}}Kr}~(\ce{^{37}Ar}).  

\subsection{Data processing}
\label{sec:dataprocessing}

The data processing consists of three stages. The first is devoted to event alignment and merging of the data files from the three ADC modules into one. The second stage is the main raw data processing of the waveforms on an event-by-event basis. It retrieves basic information about the peaks in the individual channels, identifies pulses as coincidences of these peaks and classifies their pulse type (S1/S2/noise). The third step is the post-processing tailored to the analysis phase: events are built by combining S1s and S2s based on measured charges and physical drift times. Geometry or drift time related corrections are applied later in the analysis, as detailed in Section~\ref{sec:analysis}.

\subsubsection{Pre-processing}
The first stage ensures the correct matching of the events recorded by the three ADC modules by means of a comparison of the trigger time tag changes among triggers. Misalignment among the modules can be caused by an incorrectly propagated busy state. However, an efficient offline realigning-algorithm typically restores $99.9 \, \%$ of the raw events. Once all the events are aligned, events of one dataset are merged into one ROOT file.

\subsubsection{Main processing}
\begin{figure*}[h]
\centering
\includegraphics*[width=\textwidth]{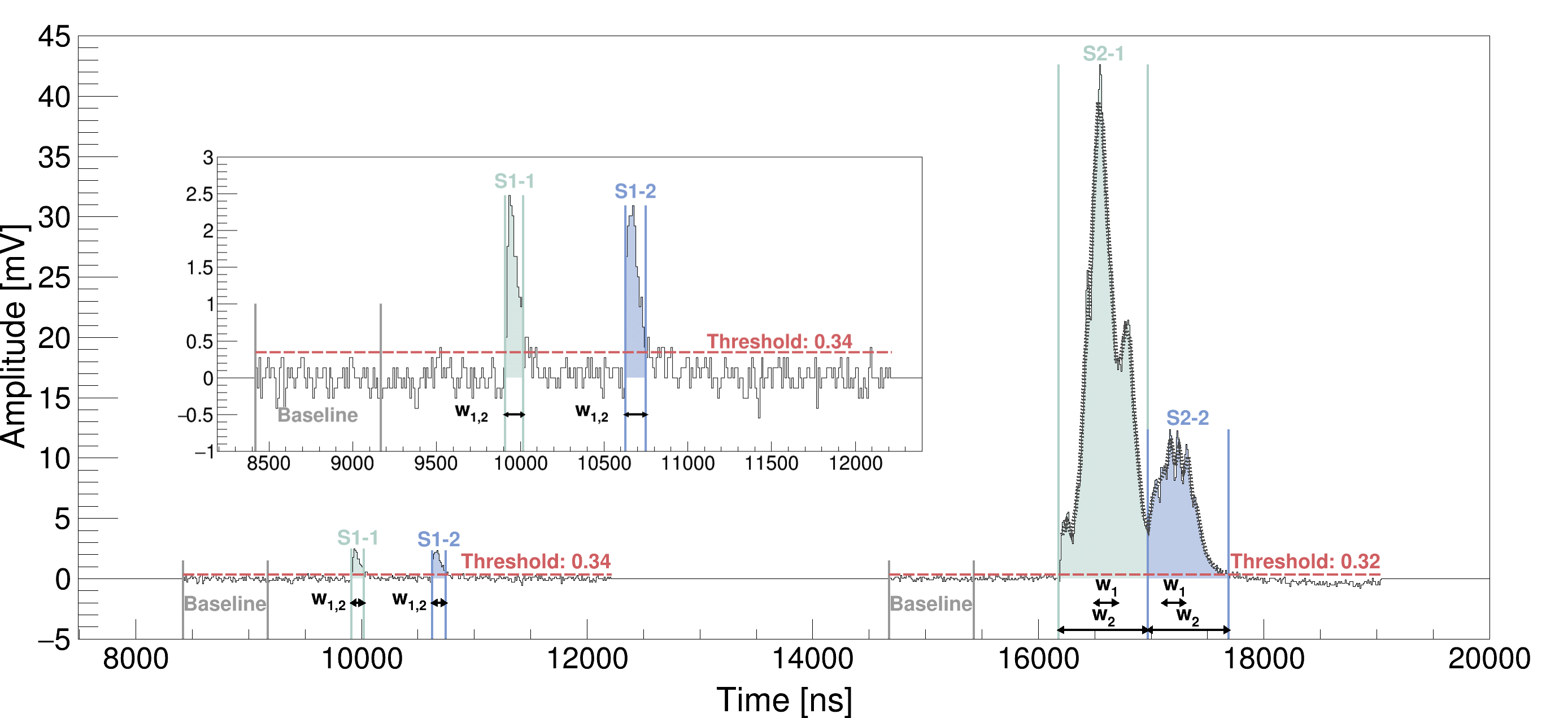}
\caption{Recorded SiPM waveform of a \ce{^{83\text{m}}Kr} event acquired with ZLE. For visualisation, we present the relevant region of the $\SI{40}{\micro s}$ DAQ window of an event from a SiPM channel located at an outer corner of the array that exhibits small S2 signals for events in the horizontal center of the TPC. The peaks are located in two good-regions surrounded by 150 backward- and forward-samples with skipped-regions in between. The integration windows are shaded and enclosed by lines. The baseline, from which the threshold is defined, is calculated individually for each good-region. While for disjointed good-regions, like here, it is based on the first half of the backward-samples, for two consecutive good-regions, the baseline is calculated within the $[60,90]\, \%$ interval between the two adjacent over-threshold regions. The S2 signals are overlaid with the moving average on which the splitting algorithm is based. While it splits the two S2s because they have both fallen below half of their individual maximum value, it is insensitive to the visible fluctuations of the S2 light. The widths $w_{1}$, $w_{2}$ used for the filters are limited to the integration window and, hence, are equal for the S1 signals. In the zoom, it is visible that over-threshold regions below 3 samples are not integrated.}
\label{fig:Wf_Proc}
\end{figure*}

Prior to the actual processing, in the ZLE case, the waveforms are reconstructed from the control words. The second stage starts with the baseline calculation of the raw waveforms and the inversion of the negative PMT signal. In the ZLE case, we account for baseline shifts by calculating the baseline for each good-region individually, shown in Figure~\ref{fig:Wf_Proc}. The threshold for peak-detection (not to be confused with the ZLE threshold for the DAQ) was chosen to be two standard deviations of the baseline distribution around zero. The peak integration limits are dynamically defined based on the variation of the baseline. Integration windows smaller than 3 samples are discarded for noise suppression.

The processor can, to a certain extent, distinguish multiple peaks that overlap. This is necessary to resolve the event topology of \ce{^{83\text{m}}Kr} with two S2~signals arriving shortly after one another (see Section~\ref{sec:kr-analysis}) and is helpful to separate afterpulses from the main peak. To reduce the sensitivity to random fluctuations, the method requires over-threshold regions of at least 6 samples. We split two peaks based on their moving average of order of 4 samples (order 2 samples for over-threshold regions $\leq 25$ samples) whenever both have fallen below half of their individual maximum value. Thus, for a Gaussian-shaped peak almost $90\%$ of its charge is integrated when cut on one side. The moving average limits the algorithm to a minimum peak separation of $\SI{40}{ns}$ or $\SI{20}{ns}$, respectively. This time-scale is sufficiently short to split even close S1~signals efficiently and is irrelevant for S2~signals for which the requirement of the contained charge dominates. 

After a peak is found, separated from others and its integration window fixed, the peak properties are extracted. The area is determined from the summed bin content of each sample, the peak position is defined as the position of the maximum sample. We also determine the width of the peak which serves as discriminator for the S1 and S2~identification. S1 and S2~signals are distinguished by width-based filters, where the filters at the $i$-th bin are defined as: 
\begin{equation}
S1_{i}\coloneqq\sum \limits_{j=i-\frac{w_{1}}{2}}^{i+\frac{w_{1}}{2}} A_{j} \quad \text{and} \quad S2_{i}\coloneqq\sum \limits_{j=i-\frac{w_{2}}{2}}^{i+\frac{w_{2}}{2}} A_{j}-S1_{i} \quad.
\end{equation}
$A_{j}$ is the baseline-subtracted signal at the $j$-th bin and $w_{1}$, $w_{2}$ are the two widths. While a maximum summation width of $w_{1}=20$ samples contains the entire S1, $w_{2}$ was chosen to be maximum 100 samples. The sums are bounded by the peak integration window to guarantee that it is not summed over a distinct close-by peak. Both filters are evaluated at~$i$ being the centre of the full width at half maximum of the peak to account for their asymmetric shape. The S2~filter will be zero for an isolated S1-like signal and much larger for an S2. We choose the ratio $S2/S1$ as discriminator and identify a peak with S2 if this ratio is $>0.2$, i.e.~if $>20 \, \%$ more charge is contained within $w_{2}$ than within $w_{1}$. This reveals the importance of good S1~splitting as we rely on the fact that S1s are contained inside~$w_{1}$. 

For the event building stage, it is essential to form physical pulses from the detected peaks. We define a pulse as a set of peaks of at least two channels in time coincidence. The coincidence window is different for S1 and S2~signals and set based on the classification of the PMT peak. 

\subsubsection{Post-processing}
The post-processing script creates high-level variables based on the described output. It selects the two highest-charge S1s and~S2s of an event with the corresponding peak properties from those pulses in which the PMT was involved. In addition, it stores the photosensor gains, reconstructs the $(x,y)$-position of each event and calculates the drift times from the delay between the S1 and S2~signals. We require a positive drift time, i.e.~the S1 must happen before the corresponding S2 and their distance must not exceed a maximum value. To reduce the contribution of high-energetic events and ensure correct S1/S2~pairing, we remove saturated signals just below the maximum input voltage of the limiting fan-in/fan-out module of $\SI{1.6}{V}$ at 11600~ADC bins or higher. The calculation of higher-level variables and the conversion from ADC bins to PE are performed at the analysis stage.

\subsection{Photosensor gain calibration}
\begin{figure}[h]
\centering
\includegraphics*[width=0.5\textwidth]{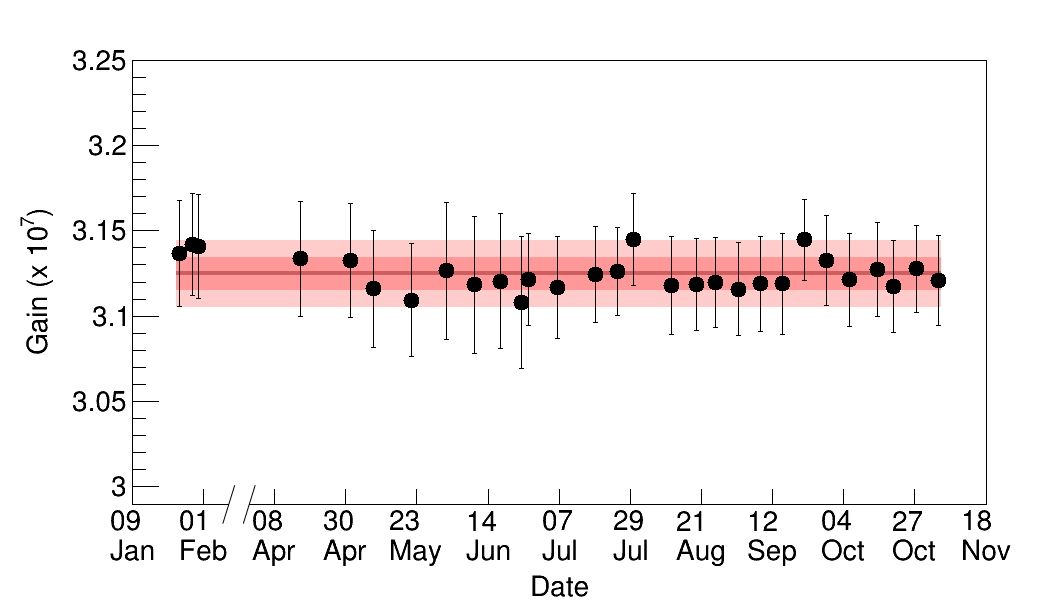}
\caption{Error-weighted mean gain (tenfold amplified) of the 16 SiPM channels over the entire data taking period in~2019. We measured $(3.12 \pm 0.01)\times 10^{7}$. The $\pm 1\sigma$~and $\pm 2\sigma$~uncertainty bands around the mean are shown in red. Data with \ce{^{83\text{m}}Kr} (\ce{^{37}Ar}) was acquired before (after) the time-axis break.}
\label{fig::gain_calibration}
\end{figure}

The gain of the photosensors was monitored and calibrated about once a week with light from an external blue LED as described in \cite{Baudis:2017xov}. The gain values are calculated following the model-independent approach proposed in \cite{Saldanha:2016}. For the bottom PMT, the measured gain during the \ce{^{37}Ar} data taking phase is $(3.76\pm0.06)\times 10^{6}$ at an operating voltage of $\SI{940}{V}$. The gain of the SiPMs is found to be uniform across the array and stable during the entire data taking. The average gain of the 16 SiPM readout channels over the data-taking period is $(3.12 \pm 0.01)\times 10^{7}$ at a bias voltage of $\SI{51.5}{V}$ including the tenfold amplification of the pre-amplifier. The given uncertainties correspond to the standard deviation of the time averaged gains. The stability of the averaged SiPM gain  is shown in Figure~\ref{fig::gain_calibration}.

\subsection{Position reconstruction and fiducial volume}
\label{sec:posrec}
\begin{figure*}[t]
\centering
\subfloat[$x$-$y$ distribution with hexagonal gate (black) and anode (white) mesh.]{
\includegraphics*[width=0.45\textwidth]{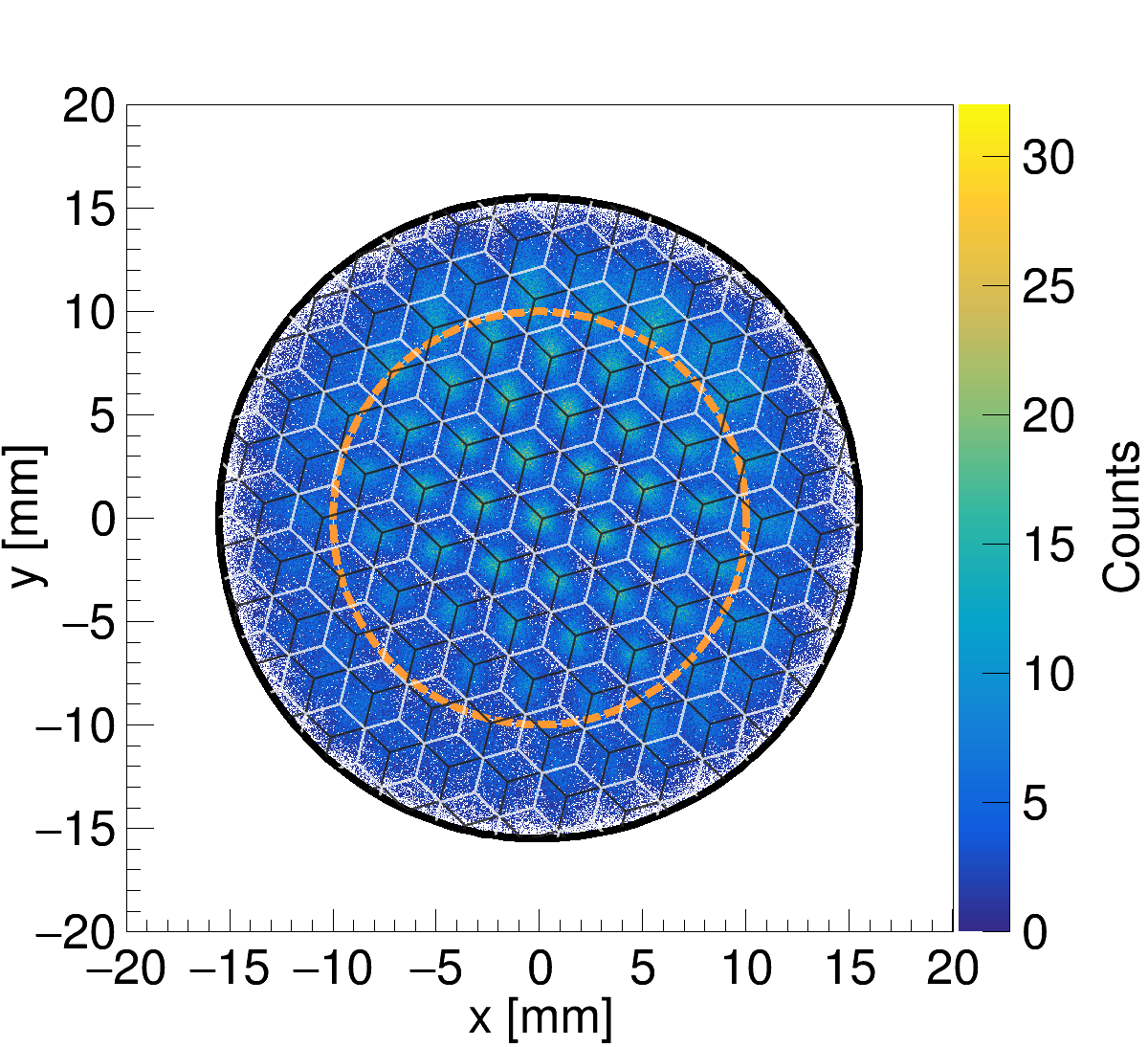}}
\hspace{1 cm}
\subfloat[$r^{2}$-$z$ distribution.]{
\includegraphics*[width=0.45\textwidth]{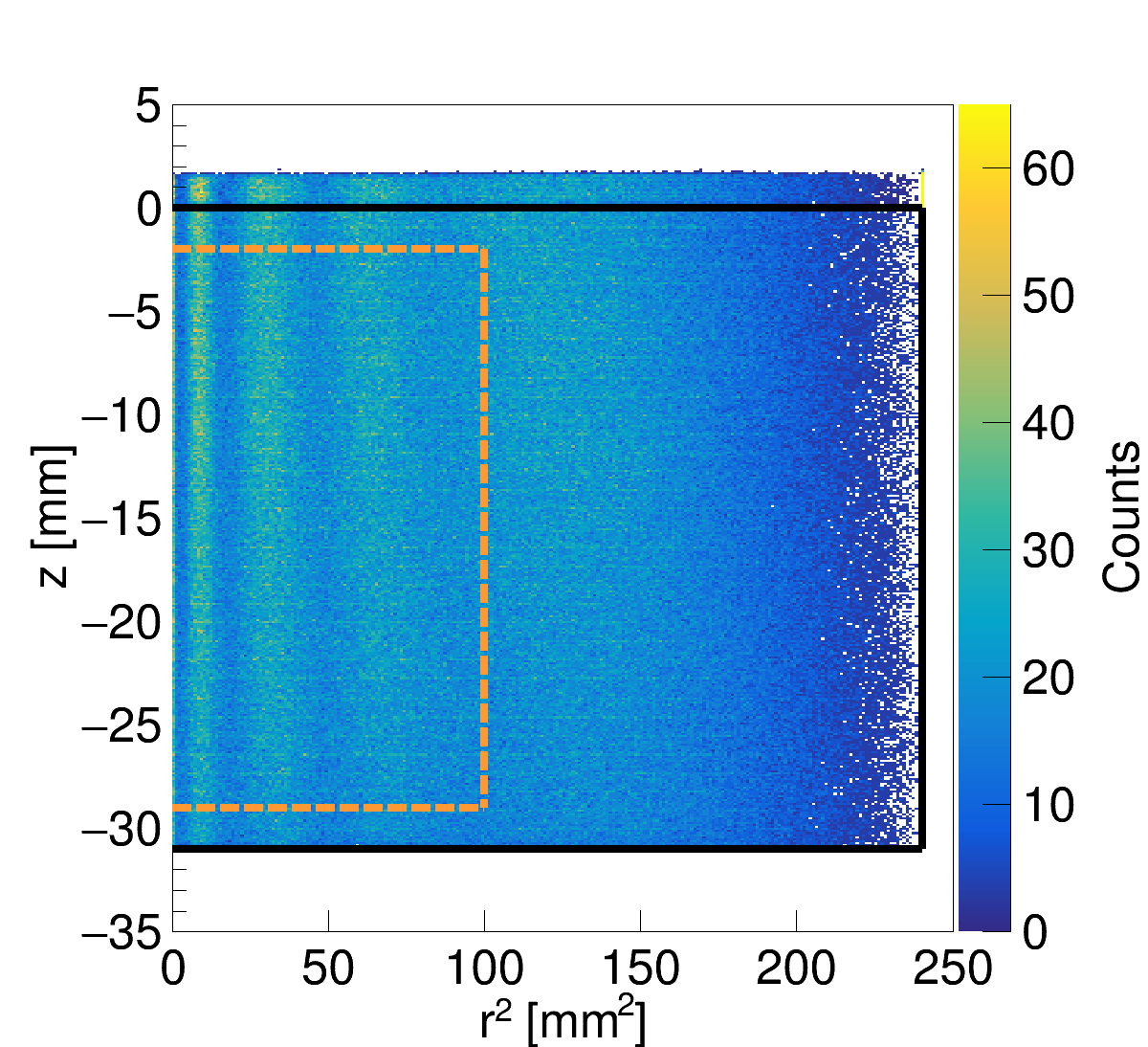}}
\caption{Event distribution from \ce{^{37}Ar} K-shell capture data. The detector volume is marked with solid black lines, the fiducial volume is represented by the dashed orange lines. On the right, the $\sim \SI{2}{mm}$ liquid level above the gate is visible.}
\label{fig:fv}
\end{figure*}
The top photosensor array allows for the reconstruction of the $(x,y)$-position of interactions. We use a simple centre-of-gravity algorithm to determine the $(x,y)$-position of an event from the light distribution among the photosensors caused by the largest S2~signal. We denote the physical position of the $i$-th SiPM in the array by $(X,Y)_{i}$, with the origin of the coordinate system being the centre-of-gravity of the array. We then obtain the (uncorrected) event position in $x$ from a geometry- and gain-weighted sum of the integrated S2~charge of each sensor\footnote{The reconstruction is performed analogously for the $y$-position.},     
\begin{equation}
x=\frac{1}{Q_{\mathrm{S2}}^{\mathrm{tot}}} \sum \limits_{i=1}^{16} X_{i} \cdot \frac{Q_{\mathrm{S2}}^{i}}{g_{i}} \quad,
\label{eq:cenofgrav}
\end{equation}
where the normalisation $Q_{\mathrm{S2}}^{\mathrm{tot}}\coloneqq\sum_{i=1}^{16} Q_{\mathrm{S2}}^{i}/g_{i}$ is the total collected S2~charge in the top array with the gain~$g_{i}$ of the $i$-th SiPM channel. 
While this approach neglects the fact that the light intensity follows an inverse-square law  with the distance from the light production site, we find that the algorithm is an adequate choice for our small-scale TPC and leads to excellent reconstruction in the TPC centre. Because of the square arrangement of the SiPMs and the nature of the centre-of-gravity approach, the resulting positions will not represent the actual circular cross-section of the TPC. In fact, while the reconstruction in the TPC centre is very good, events at the TPC boundary feature a square-shaped bias. For this reason, the square is scaled down to unit side length with a constant, $x_{\mathrm{scal}}=x/c$, and then mapped onto a unit circle that is scaled back,
\begin{equation}
x_{\mathrm{map}}=c \cdot x_{\mathrm{scal}} \cdot \sqrt{1-\frac{y^2_{\mathrm{scal}}}{2}} \quad.
\end{equation}
While these mapped positions are centred around the origin of the coordinate system due to the gain correction in Equation~\ref{eq:cenofgrav}, they still feature a radial bias because of the solid angle seen by the individual photosensors. Using a data-driven approach, we correct for this effect by comparing the positions of the spots visible in the $(x,y)$-distribution with the known gate mesh junctions. In Figure~\ref{fig:fv}, left, we show the $(x,y)$-distribution from \ce{^{37}Ar} data overlaid by the drawing of the meshes. The spots occur due to focusing of drifting electrons to the high-field regions around the junctions of the grounded hexagonal gate mesh. This limits the accuracy of the $(x,y)$~event position reconstruction to roughly half the distance between the junctions, which is $\sim \SI{1.5}{mm}$.

Up to a radius of $\sim \SI{10}{mm}$ from the TPC centre, the correction function is fairly linear, i.e.~the correction can be performed with a scaling factor, whereas for larger radii it diverges. This can be described by a projection of a sphere onto a plane as:
\begin{equation}
x_{\mathrm{corr}}=\frac{d_\mathrm{TPC} \cdot \left(\frac{1}{2}-\frac{1}{\pi} \cdot \arccos{\frac{r_{\mathrm{map}}}{r_{\mathrm{map,max}}}}\right)}{r_{\mathrm{map}}}\cdot x_{\mathrm{map}} \quad,
\end{equation}     
where $r_{\mathrm{map}}=\sqrt{x_{\mathrm{map}}^2+y_{\mathrm{map}}^2}$ is the mapped radius of the interaction site with $r_{\mathrm{map,max}}$ being the boundary of the TPC after the mapping and $d_\mathrm{TPC}=\SI{31}{mm}$ the true diameter of the TPC. The linear region of this projection defines our radial fiducial volume cut to be $r<\SI{10}{mm}$.
The $z$-position of an interaction is obtained from the time difference between the S1 and S2~signals. We extract the electron drift speed for a given electric field from the clearly visible position of the cathode and the gate in the drift time histogram and their spatial separation. The procedure, together with a drift speed measurement at different fields in Xurich~II, are detailed in~\cite{Baudis:2017xov}. The drift speeds that we find here are in agreement with this former measurement and are in the range $\SIrange{1.5}{2.0}{\si{mm/\micro s}}$.

To avoid high and potentially non-uniform electric field regions around the electrodes, we select only events within $z \in [-29,-2] \, \si{mm} $, where $\SI{0}{mm}$ is the position of the gate and $-\SI{31}{mm}$ is the position of the cathode. The fiducial region defined by these boundaries is shown in Figure~\ref{fig:fv} and contains a xenon mass of $\sim \SI{24.5}{g}$.

\section{Data analysis}
\label{sec:analysis}

In this section, we describe the data selection, the applied quality cuts and the signal corrections. We require correctly paired S1 and S2~events for both calibration sources. The focus is on the \ce{^{37}Ar} source, since it is a rather new approach to calibrate the low-energy region, while calibrations with \ce{^{83\text{m}}Kr} are well established. In this work, we show \ce{^{37}Ar} data from the K-shell capture, with a line at $\SI{2.82}{keV}$ only.

\subsection{Energy calibration with \ce{^{37}Ar}}
\subsubsection{Data selection and quality cuts}
The data used for the calibration was taken between late-May and early-July 2019 while the campaign for the half-life measurement lasted until mid-October 2019. 

\begin{figure*}[t]
\centering
\subfloat[Area fraction top cut to remove gas events with a fraction of $\SIrange{0.3}{0.5}{}$. Besides the K-shell population on the right we also select parts of the L-shell population below 600 PE S2b and the single photoelectron region that are both mostly located in S2-only space. These are removed by an energy cut.]{
\includegraphics*[width=0.45\textwidth]{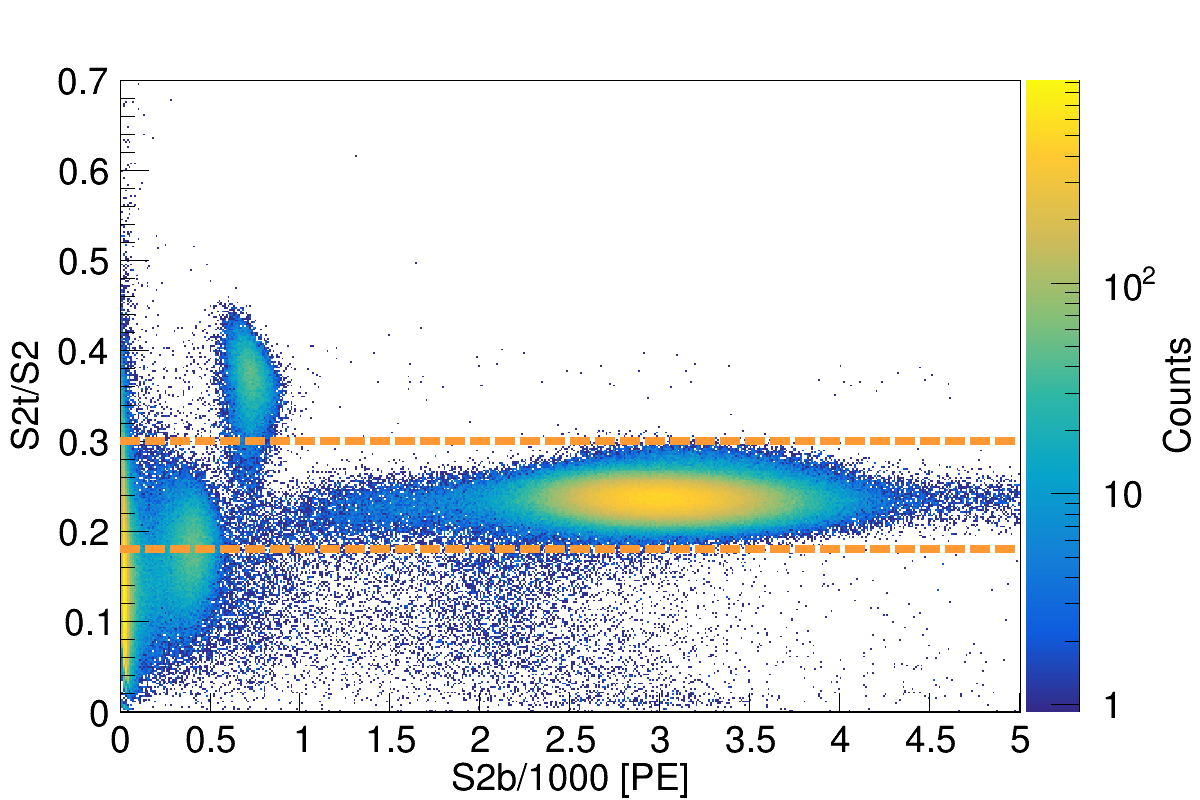}}
\hspace{1 cm}
\subfloat[S2 width cut to remove accidental coincidences and remaining gas events (see text). The drift region defined by the cathode and the gate are marked by solid black lines.]{
\includegraphics*[width=0.45\textwidth]{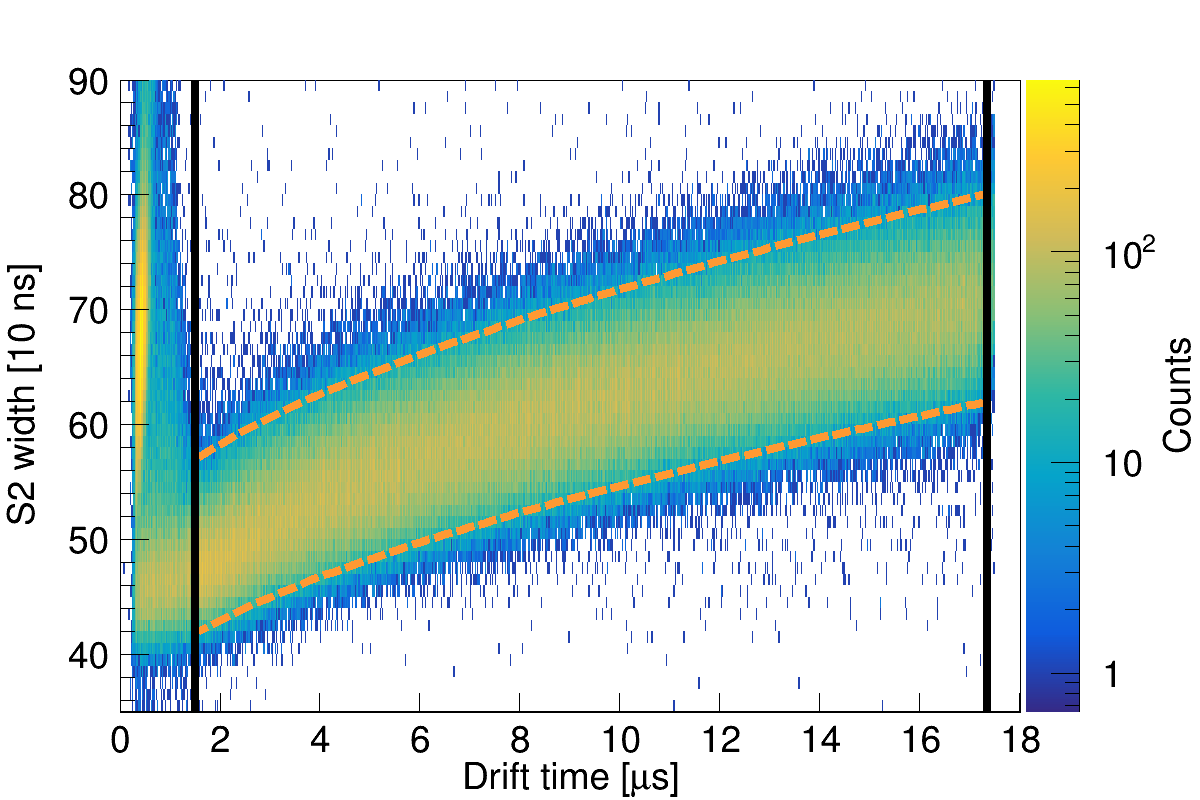}}
\caption{Data quality cuts. The selected region is enclosed by the dashed orange lines.}
\label{fig:Cuts}
\end{figure*}

As described at the end of Section~\ref{sec:dataprocessing}, the post-processing accounts for physical drift times and removes saturated signals. Because of the low \ce{^{37}Ar} signal rate in the detector volume, $\mathcal{O}(\SI{10}{Hz})$, we expect a negligible pile-up rate for a DAQ window of $\SI{60}{\micro s}$. This includes pile-up with background events with a rate of the same order of magnitude. Furthermore, we select only single S2s, for we do not expect multiple scatters from a single \ce{^{37}Ar} decay. To ensure correct S1 and S2~matching, we  require the S2 signal to contain more charge than the S1. In addition, we apply the fiducial volume cut described in Section~\ref{sec:posrec} to remove wall and gas events, regions with an inhomogeneous electric field, and those outside of the linear region of the $(x,y)$-position reconstruction algorithm. To efficiently remove events in the gas phase that yield a large S2 in the top array compared to the total S2, we apply a so-called area fraction top cut shown in Figure~\ref{fig:Cuts}, left.
To remove the lower energy lines from the analysis, we apply a cut on the energy, namely in the centre of the $5\sigma$ boundaries of the L-and the K-shell populations in the total S1+S2~space. Since accidental coincidences have by definition a random drift time, they can be efficiently removed with a cut on a variable with a strong drift time dependence. The longitudinal diffusion of the electron cloud during the drift leads to such a dependence of the S2~width, as shown in Figure~\ref{fig:Cuts}, right. This distribution follows an empirical function with three parameters:
\begin{equation}
\mathrm{S2_{width}} (t_{\mathrm{drift}}) = P_0 + P_1 \cdot t_{\mathrm{drift}} + \sqrt{P_2 \cdot t_{\mathrm{drift}}} \quad.
\end{equation}
Here, the width corresponds to the full width at tenth maximum. The upper and lower cuts are defined as follows: the histogram is sliced in drift time bins, and the fit is performed through the $97.7\,\%$ and $2.3\,\%$ quantiles of those bins. We thus select the $\pm 2 \sigma$ region around the mean. For \ce{^{37}Ar} data, $5.6 \, \%$ of the recorded events that generate an S2 signal pass all data quality cuts.

\subsubsection{Corrections}
\begin{figure*}[t]
\centering
\subfloat[Electron lifetime correction fit with relative residuals. The electron lifetime from the combined data at $\SI{968}{V/cm}$ is $(\SI{124.2}{} \pm \SI{0.7}{})\si{\, \micro s}$. The error is the error on the fit parameter and purely statistical.]{
\includegraphics*[width=0.45\textwidth]{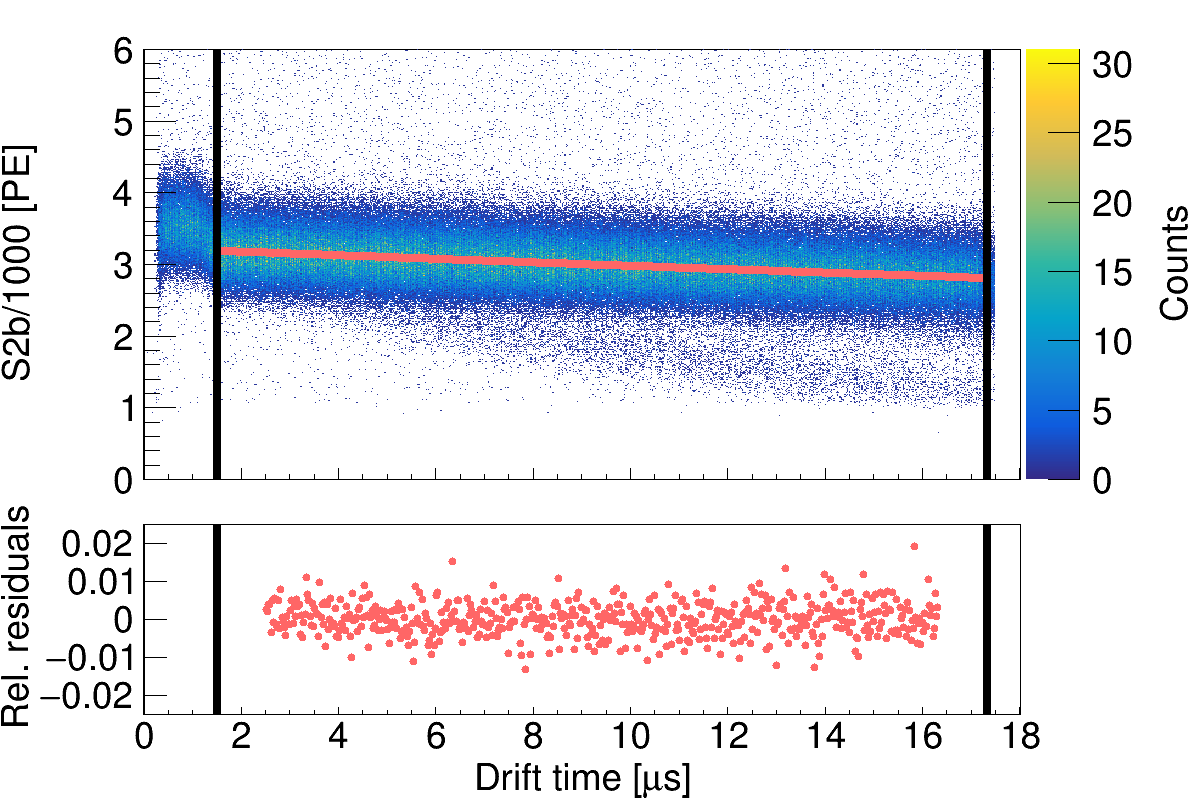}}
\hspace{1 cm}
\subfloat[S1b correction fit with relative residuals.]{
\includegraphics*[width=0.45\textwidth]{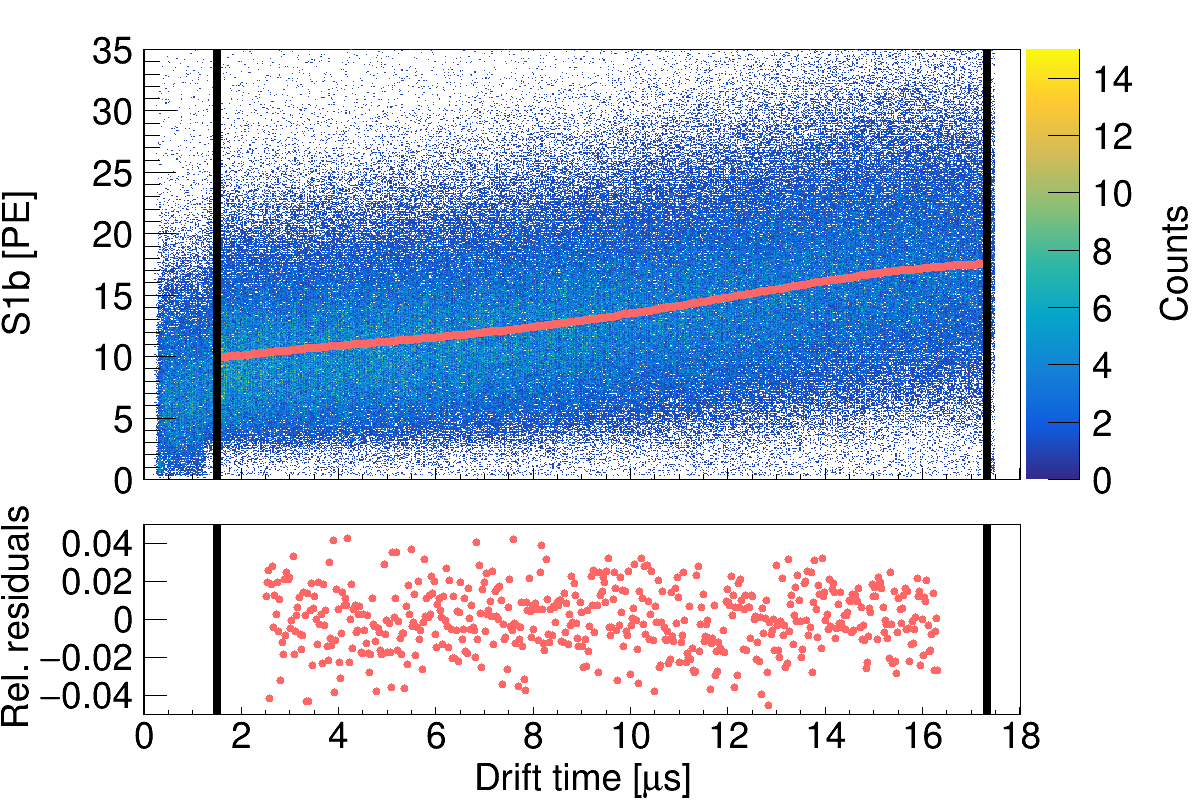}}
\caption{Corrections for systematic detector effects. The drift region defined by the cathode and the gate is marked by solid black lines.}
\label{fig:Corrections}
\end{figure*}

\begin{figure}[h]
\centering
\includegraphics*[width=0.5\textwidth]{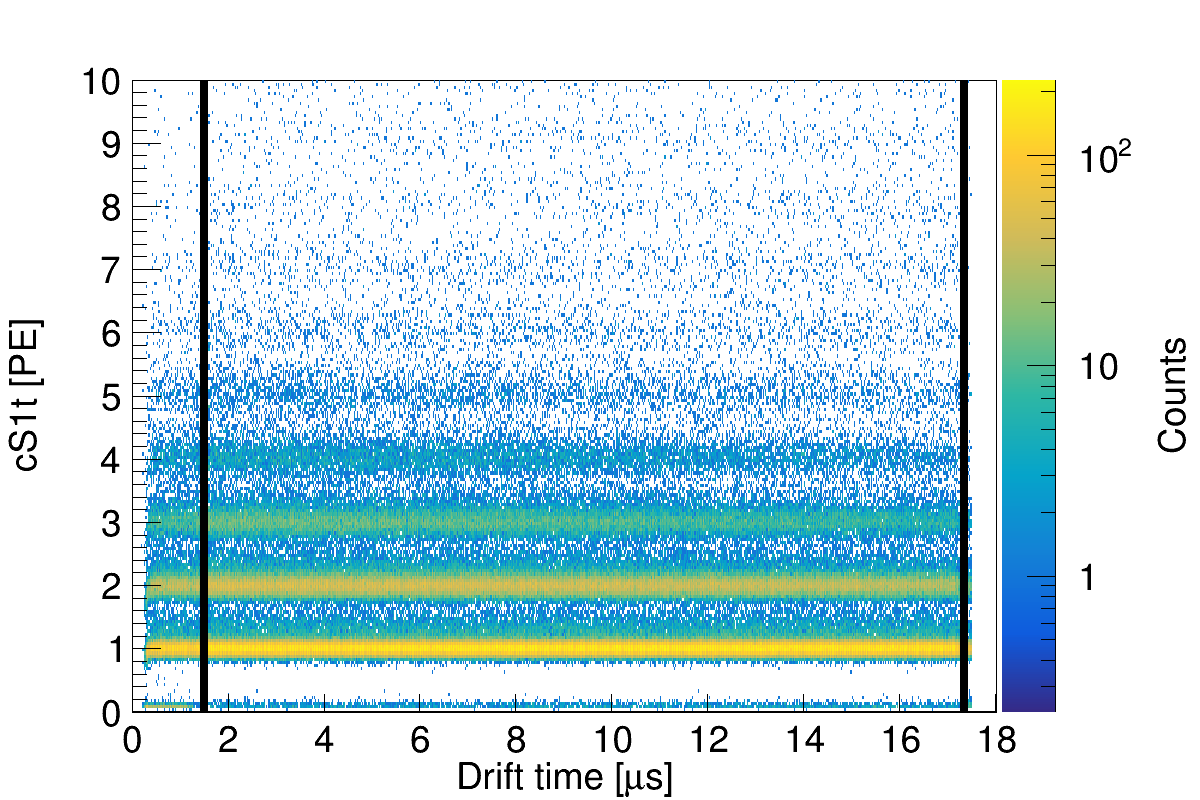}
\caption{Corrected S1t vs.~drift time histogram with the single photoelectron band structure.}
\label{fig:Corrections2}
\end{figure}

As a result of the recombination of the drifting free electrons with electronegative impurities in LXe, the S2~charge degrades with drift time. The free electron lifetime $\tau_{\mathrm{e}}$, defined as the decay constant in an exponential decay law, is a measure of the LXe purity. Being a property of the LXe, the drift time dependence of the S2 signal of any channel can be used to correct for this systematic effect. We base our correction on the bottom PMT as it collects more S2 light than the top array. To this end, we fit the $50\,\%$ quantiles of the 2-dimensional S2b\footnote{In the following b refers to the PMT channel (\emph{bottom}) and t to the combined SiPM channels (\emph{top}).} versus drift time histogram with an exponentially decreasing function (see Figure~\ref{fig:Corrections}, left). We thus obtain the corrected S2~charge by multiplying with its inverse, i.e.~we scale according to the S2 at zero drift time:
\begin{equation}
\mathrm{cS2}=\mathrm{S2} \cdot \exp{\left( t_{\mathrm{drift}}/\tau_{\mathrm{e}} \right)} \quad.
\end{equation}  
The stable detector conditions during the time in which the calibration data was taken resulted in an electron lifetime of $\SIrange{85}{187}{\micro s}$ without showing any abrupt changes. While we correct for the electron lifetime on a daily basis, in Figure~\ref{fig:Corrections}, left, we show for simplicity the combined data at $\SI{968}{V/cm}$ drift field with an averaged electron life time.     
The LCE depends on the geometry and the optical properties of the TPC. In particular, the collected S1~light varies with the depth of the event. We use a data-driven approach and fit the $50\,\%$ quantiles of the 2-dimensional S1b versus drift time histogram with a 4th order polynomial, as shown in Figure~\ref{fig:Corrections}, right. The corrected signal $\mathrm{cS1b}$ is obtained by scaling around the drift time corresponding to the centre of the TPC. 
Thanks to the excellent single photoelectron resolution of SiPMs, the S1t histogram features a band-like structure, as visible in Figure~\ref{fig:Corrections2}. Since the small light signal seen by the SiPMs does not show any drift time dependency within a single band, no data-based drift-time dependent correction could be performed for S1t. Because of this discreteness, we would need to know which events changed to the next lower band after a certain drift time. However, these bands help us to fix the charge scale in PE. To that end, Gaussian fits on the seven lowest bands are performed in the S1t~histogram and then linearly mapped onto the expected charge in PE. Figure~\ref{fig:Corrections2} shows the corrected result.
All the described corrections are performed on data within the fiducial volume only. For simplicity, we do not apply $(x,y)$-dependent corrections.

\subsection{Energy calibration with \ce{^{83\text{m}}Kr}}
\label{sec:kr-analysis}

A common technique for the energy calibration of a xenon dual-phase TPC is to introduce \ce{^{83\text{m}}Kr} into the system~\cite{Manalaysay:2009yq,Aprile:2017aty,LUX:2017,Hannen:2011}. This source with $T_{1/2}=\SI{1.83 (2)}{h}$ offers a homogeneous distribution of events in the target volume with a line at $\SI{32.15}{keV}$ followed by a line at $\SI{9.41}{keV}$ with $T_{1/2}=\SI{155.1 (12)}{ns}$ for the intermediate state~\cite{McCutchan:2015}.  

\subsubsection{Data selection and quality cuts}

The selection of \ce{^{83\text{m}}Kr} events is performed by exploiting the double-S1 and double-S2~signal topology as a result of the intermediate $\SI{9.41}{keV}$ level. In the following, we only consider split S1 and S2-pairs and not the combined $\SI{41.56}{keV}$ line. That is, in the PMT channel, we require  exactly two S1s with S2s and the S1s must contain less charge than their paired S2s. In a next step, a cut on the delay time of the PMT signals is applied. We expect \ce{^{83\text{m}}Kr} events to have the same positive delay $\Delta t$ between the two S1~signals as between the two S2~signals and select $\pm 2\sigma$ around the centred mean in the $\Delta t_{\mathrm{S1}}-\Delta t_{\mathrm{S2}}$ histogram. As for \ce{^{37}Ar}, we use the fiducial volume cut from Section~\ref{sec:posrec}, an area fraction top cut, an energy cut to separate the lines, as well as the S2 width cut. For \ce{^{83\text{m}}Kr} data, $1.9 \, \%$ of the recorded events that generate an S2 signal pass all data quality cuts. 

\subsubsection{Corrections}

We apply the same corrections on the \ce{^{83\text{m}}Kr} lines as described above for \ce{^{37}Ar}. Due to the higher energy of the source, and thus the higher S1~signal  seen by the top SiPMs, we are  able to correct S1t with a 4th order polynomial showing qualitatively the opposite drift time dependency compared to the S1b signal (cf.~Figure~\ref{fig:Corrections}, right).

\section{Results and Discussion}
\label{sec:results}

\begin{figure*}[t]
\centering
\subfloat[$\SI{2.82}{keV}$ line of \ce{^{37}Ar}. The charge and light anti-correlation is moderated at this low energy.]{
\includegraphics*[width=0.45\textwidth]{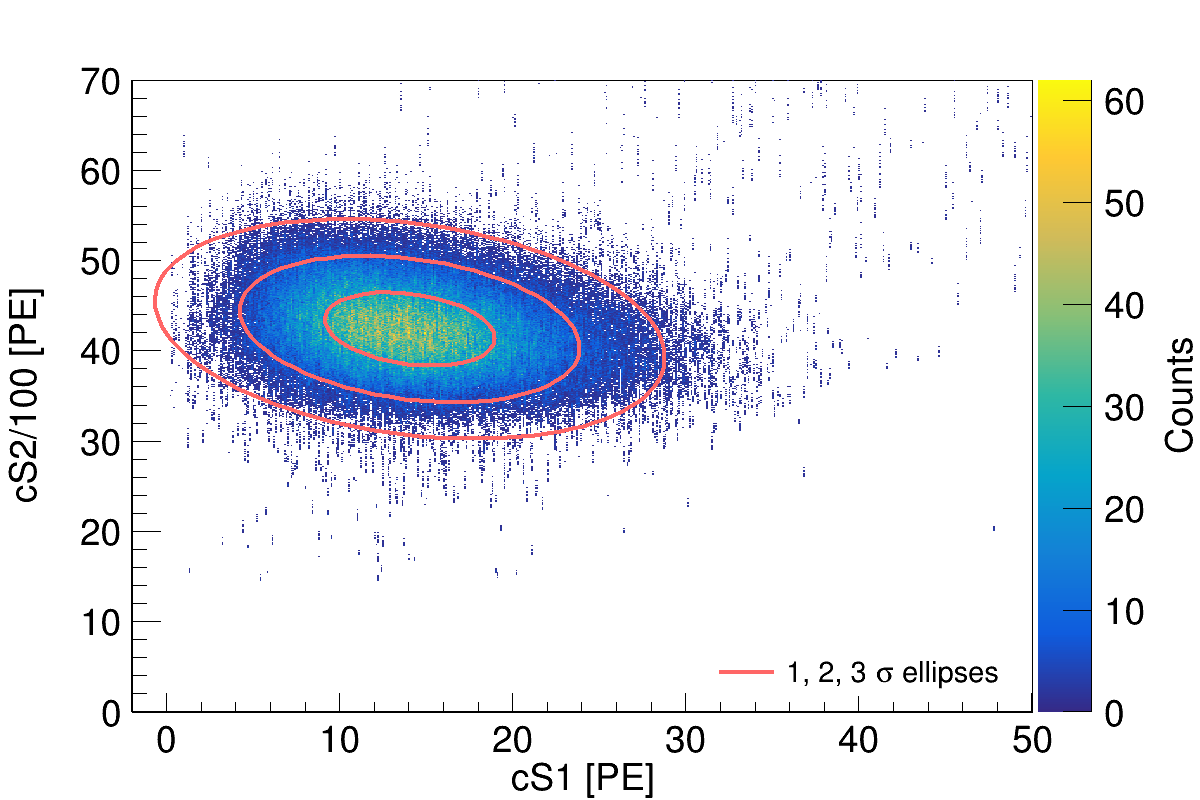}}
\hspace{1 cm}
\subfloat[$\SI{9.41}{keV}$ and $\SI{32.15}{keV}$ line of \ce{^{83\text{m}}Kr}. The anti-correlation is clearly visible. The color scale is bounded above for visualisation.]{
\includegraphics*[width=0.45\textwidth]{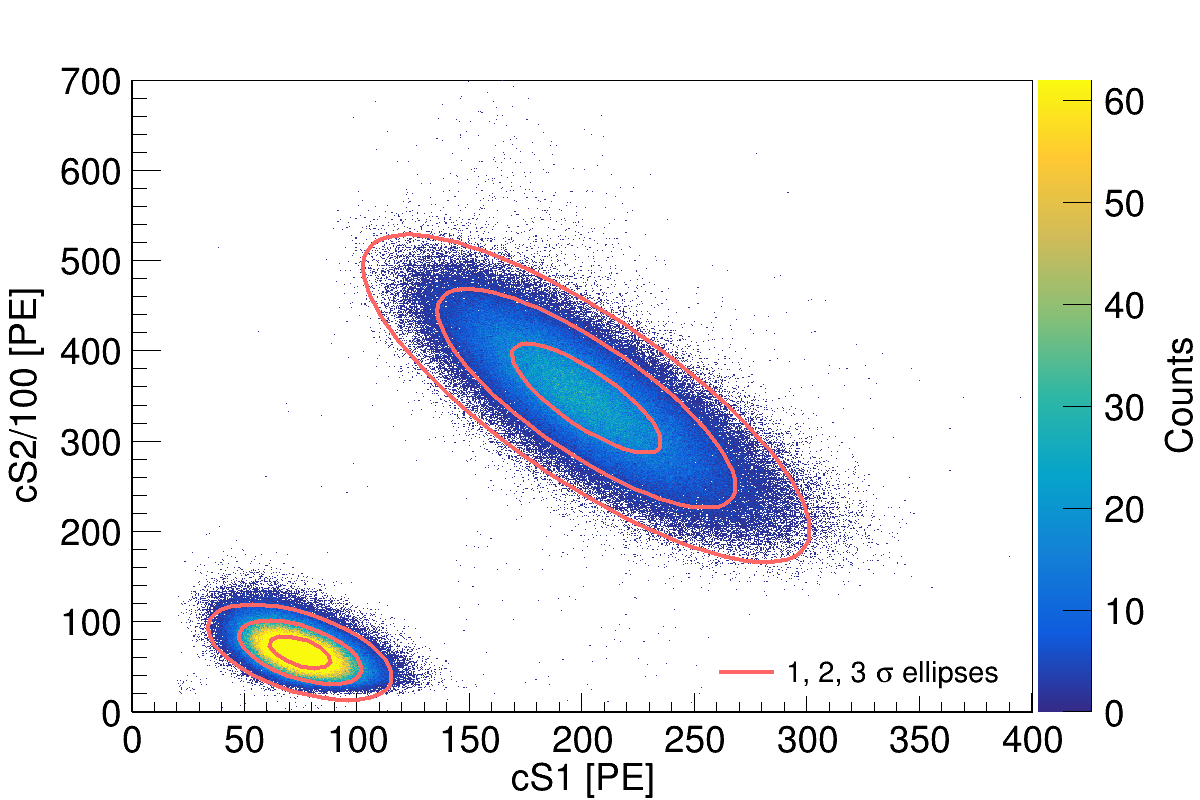}}
\caption{Low energy populations of both sources in total S2 versus total S1 space showing a charge and light anti-correlation. The data was acquired at a drift field of $\SI{968}{V/cm}$. The ellipses show the $1, \, 2, \, 3 \, \sigma$ regions of the distribution.}
\label{fig:ellipses}
\end{figure*}

After applying the described corrections, we show in Figure~\ref{fig:ellipses} the localised populations from the $\SI{2.82}{keV}$ line of \ce{^{37}Ar} (left) together with the two lines of \ce{^{83\text{m}}Kr} (right) in total S2 versus total S1~space. The anti-correlation between ionisation and scintillation signals is evident, however less pronounced at the low energy of \ce{^{37}Ar} due to the small S1 signals.

\begin{figure}[h]
\centering
\includegraphics*[width=0.45\textwidth]{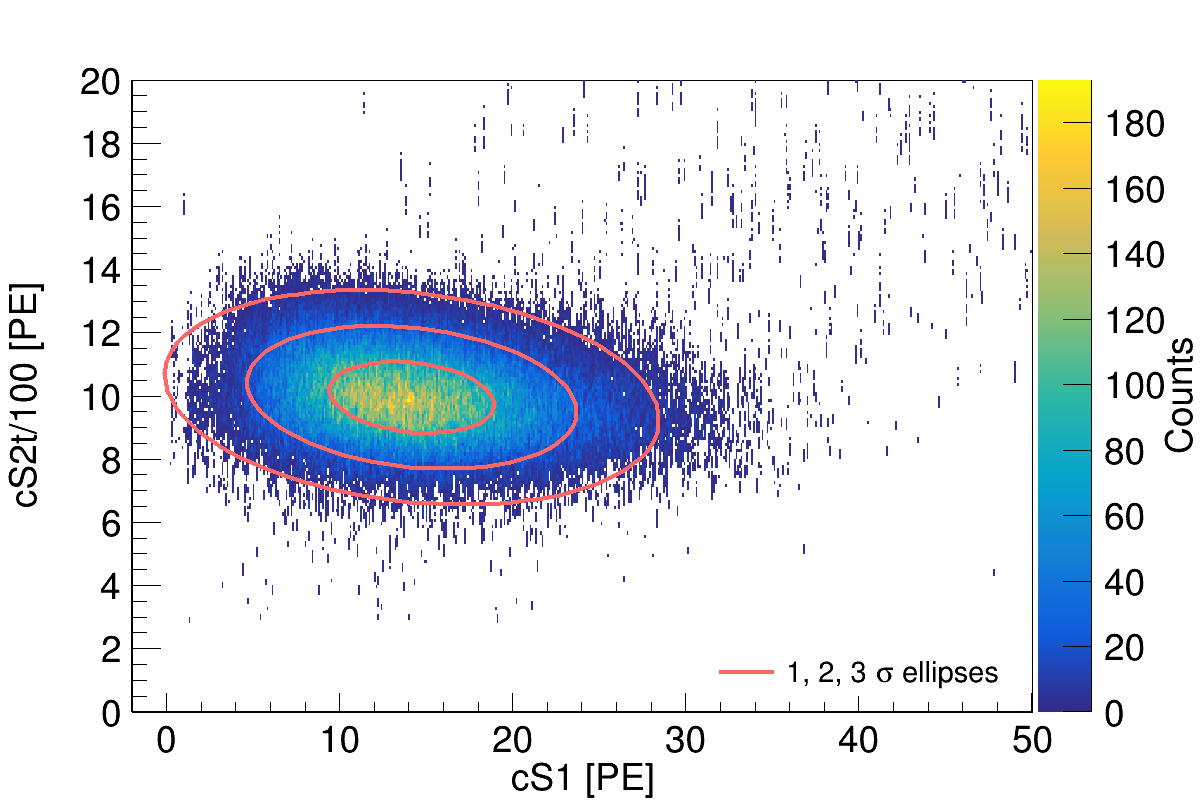}
\caption{$\SI{2.82}{keV}$ line of \ce{^{37}Ar} in S2 top versus total S1 space at $\SI{968}{V/cm}$. The ellipses show the $1, \, 2, \, 3 \, \sigma$ regions of the distribution. The SiPM top array collects $23.5 \, \%$ of the total detected S2 light (cf.~Figure~\ref{fig:ellipses}, left).}
\label{fig:top_ellipse}
\end{figure}

From the mean of the \ce{^{37}Ar}~distribution, we deduce a light yield of $(\SI{4.985}{} \pm \SI{0.003}{}) \SI{}{\, PE/keV}$ and a charge yield of $(\SI{1504.0}{} \pm \SI{0.2}{}) \SI{}{\, PE/keV}$ for $\SI{2.82}{keV}$ at $\SI{968}{V/cm}$ drift field. For \ce{^{83\text{m}}Kr}, at the same drift field, we obtain a light yield of $(\SI{7.941}{} \pm \SI{0.003}{}) \SI{}{\, PE/keV}$ and a charge yield of $(\SI{702.0}{} \pm \SI{0.3}{}) \SI{}{\, PE/keV}$ at $\SI{9.4}{keV}$, and $(\SI{6.284}{} \pm \SI{0.002}{}) \SI{}{\, PE/keV}$ and $(\SI{1080.1}{} \pm \SI{0.3}{}) \SI{}{\, PE/keV}$, respectively at $\SI{32.1}{keV}$. 
The uncertainties reflect the statistical errors on the mean of the yields. The discussion of the systematics, in particular the limitation of the \ce{^{83\text{m}}Kr} charge yield precision, follows below. In Figure~\ref{fig:top_ellipse}, we present the $\SI{2.82}{keV}$ population of \ce{^{37}Ar} for the S2 light of the top array only. From the comparison with Figure~\ref{fig:ellipses} (left), we deduce that $23.5 \%$ of the total detected S2 light is collected by the SiPMs.

\begin{figure}[h]
\centering
\includegraphics*[width=0.5\textwidth]{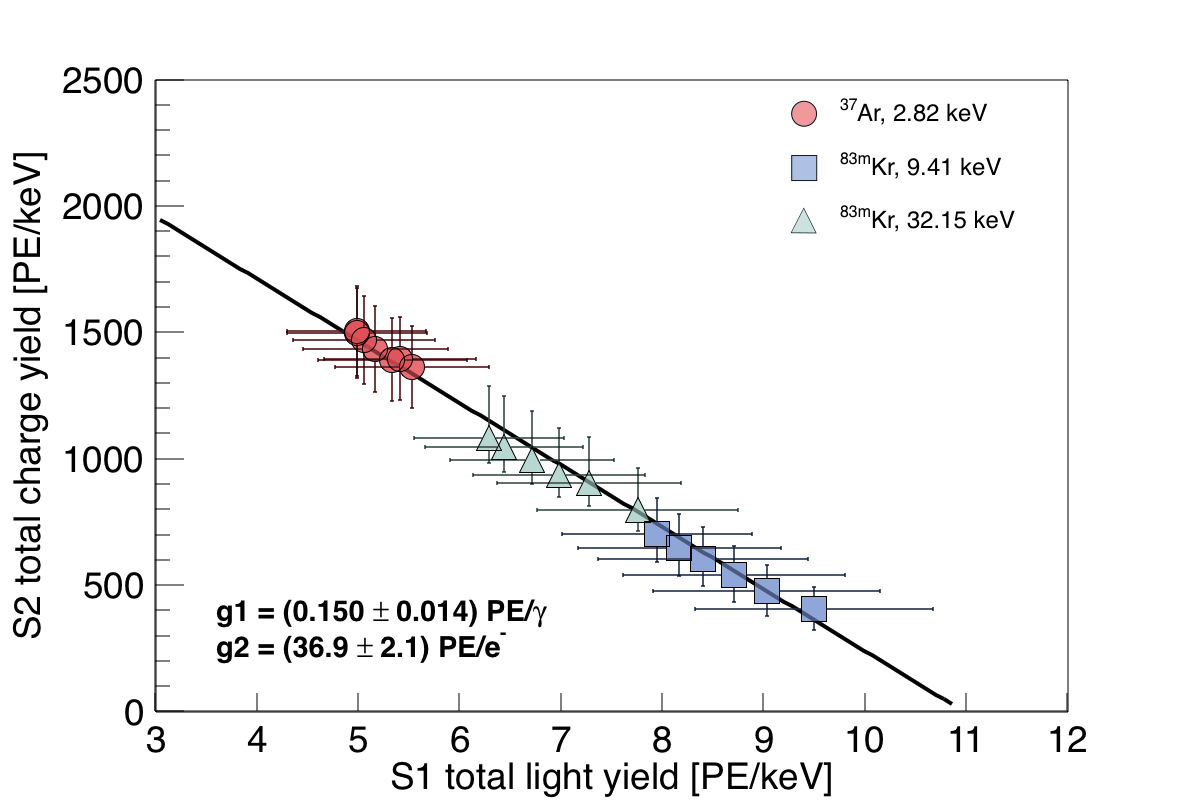}
\caption{Anti-correlation of ionisation and scintillation signals at different drift fields in the range $\SIrange{80}{968}{V/cm}$ for \ce{^{37}Ar} and $\SIrange{194}{968}{V/cm}$ for \ce{^{83\text{m}}Kr}, respectively. The axis intersections yield the detector-specific gains $g1$ and $g2$ from the combined energy scale.}
\label{fig:Doke}
\end{figure}

Calculating the total light and charge yield for data at different drift fields yields the so-called \emph{Doke plot}~\cite{Doke:2002}, shown in Figure~\ref{fig:Doke}. The light and charge yield measurements of the three lines from \ce{^{37}Ar} and \ce{^{83\text{m}}Kr} feature a clear anti-correlated linear dependence. The \ce{^{37}Ar} points are less spread than those from  \ce{^{83\text{m}}Kr} because the drift field dependence of the ionisation and scintillation yield decreases rapidly below $\sim \SI{10}{keV}$ ER energy~\cite{Goetzke:2017}. The errors include statistical uncertainties, as well as systematic uncertainties of the PMT gain and of the peak-splitting method described in Section~\ref{sec:dataprocessing}. The condition on the contained charge explains the systematic bias of the $\SI{32.15}{keV}$ points towards lower charge yield and is the origin of the asymmetric vertical error bars assigned to \ce{^{83\text{m}}Kr}. However, this is not relevant for the light yield as the S1 delay time is sufficiently long to provide complete S1 separation for any split S2 signal.  

\begin{figure}[h]
\centering
\includegraphics*[width=0.5\textwidth]{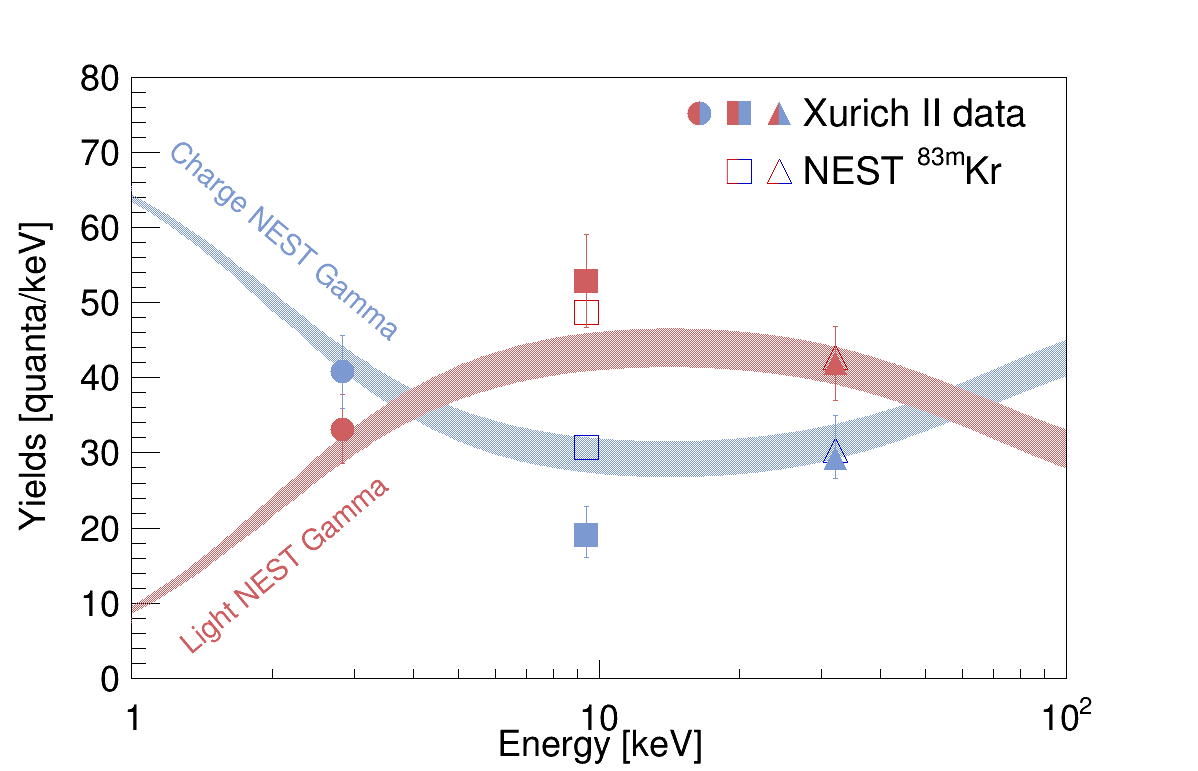}
\caption{Comparison of the obtained charge (blue) and light (red) yields from \ce{^{37}Ar} and \ce{^{83\text{m}}Kr} lines at $\SI{968}{V/cm}$ to the NEST~v2.0 gamma and \ce{^{83\text{m}}Kr} model. The uncertainty bands reflect the intrinsic systematics of the gamma model.}
\label{fig:NEST}
\end{figure}

The detector-specific gains $g1$ and $g2$ are defined as
\begin{equation}
\label{eq:g1g2}
g1 \coloneqq S1/N_{\gamma} \quad \text{and} \quad g2 \coloneqq S2/N_{e^{-}} \quad,
\end{equation} 
where $N_{\gamma}$ is the total number of photons produced by de-excitation or recombination and $N_{e^{-}}$ is the total number of electrons producing the S2 signal. With $W=(\SI{13.7}{} \pm \SI{0.2}{})\SI{}{\, eV}$ being the mean energy required to produce an excited or ionised xenon atom~\cite{DahlThesis}, the combined energy scale of an event is
\begin{equation}
E=W \left( N_{\gamma}+N_{e^{-}} \right)=W \left( \frac{S1}{g1}+\frac{S2}{g2} \right) \quad,
\end{equation}
where we substituted the number of quanta for the definitions in Equation~\ref{eq:g1g2}. 
From the axis intersections of the linear regression in Figure~\ref{fig:Doke}, we obtain $g1=(\SI{0.150}{} \pm \SI{0.014}{})\SI{}{\, PE/\gamma}$ and $g2=(\SI{36.9}{} \pm \SI{2.1}{})\SI{}{\, PE/e^{-}}$. The combined energy resolution $\sigma/E$ from S1 and S2 signals of the $\SI{2.82}{keV}$  \ce{^{37}Ar} line is $(\SI{17.2}{} \pm \SI{0.1}{}) \, \%$ at $\SI{968}{V/cm}$ and stays constant down to $\SI{600}{V/cm}$. Below this value, the resolution degrades  slowly to $(\SI{18.0}{} \pm \SI{0.1}{}) \, \%$ at the lowest measured field. 
The resolution at the highest energy of $\SI{32.15}{keV}$ is $(\SI{5.80}{} \pm \SI{0.01}{}) \, \%$ which is consistent with the one we previously measured with the TPC with two PMTs~\cite{Baudis:2017xov}.
The yields of all lines at $\SI{968}{V/cm}$ are compared to the Noble Element Simulation Technique (NEST)~v2.0~\cite{Brodsky:2019} model predictions in Figure~\ref{fig:NEST}. Both charge and light yield at $\SI{2.82}{keV}$ and $\SI{32.15}{keV}$ agree well with the gamma model. In addition, we show the predictions for the two lines of \ce{^{83\text{m}}Kr}. While we see good agreement at $\SI{32.15}{keV}$, we find a lower charge yield than predicted at $\SI{9.41}{keV}$. Due to anti-correlation, the light yield at the latter energy is higher than predicted but consistent with NEST~v2.0 \ce{^{83\text{m}}Kr} within the error.  

\begin{figure}[h]
\centering
\includegraphics*[width=0.5\textwidth]{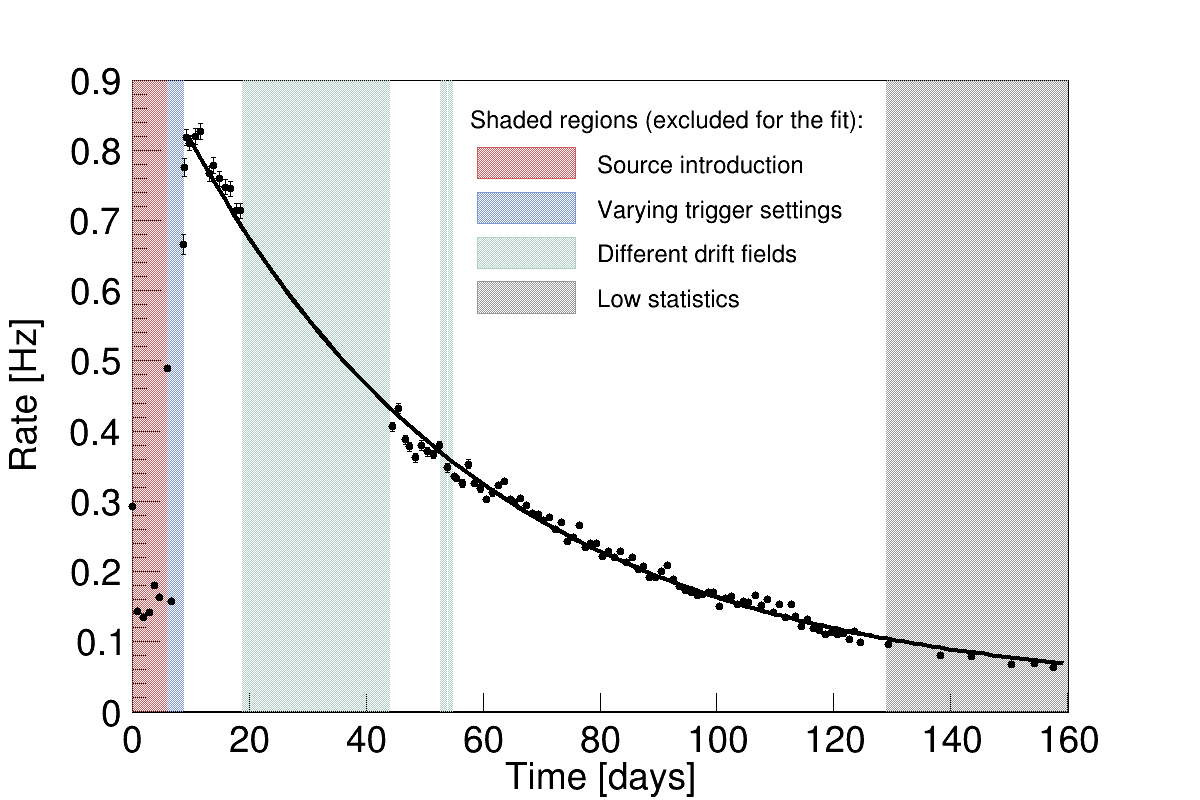}
\caption{Activity evolution of the selected $\SI{2.82}{keV}$ data of \ce{^{37}Ar} over more than 4 half-lives with a one-day binning starting from the time at which the first ampule was broken. The fit was performed on $\SI{968}{V/cm}$ data only, because at other drift fields the data features systematic deviations caused by the varying efficiency of the applied data-driven cuts and corrections. The errors on the individual data points are Poisson-like. We obtain a half-life of $(\SI{35.7}{} \pm \SI{0.4}{})\si{\, \day}$.}
\label{fig:ar_half-life}
\end{figure}

Data with the \ce{^{37}Ar} source was acquired over a period of more than 4 half-lives. Selecting the data as described in Section~\ref{sec:analysis} in daily slices, we show in Figure~\ref{fig:ar_half-life} the evolution of the activity of the $\SI{2.82}{keV}$ line. For simplicity, we assume that all the systematics and applied cuts are stable over the time of the measurement. The exponential fit yields a half-life of $(\SI{35.7}{} \pm \SI{0.4}{})\si{\, \day}$, where the most recent literature value is $(\SI{35.011}{} \pm \SI{0.019}{})\si{\, \day}$ from a weighted average of four measurements~\cite{Cameron:2012}. The given error is the error of the decay parameter of the least squares fit and does not include systematic uncertainties. However, within the scope of this work, we can be confident that we indeed selected events from \ce{^{37}Ar} decays.

For \ce{^{83\text{m}}Kr}, exponential fits of the S1 and S2~delay histograms yield $(\SI{161}{} \pm \SI{5}{})\si{\, ns}$ for the half-life of the intermediate state, in agreement with the literature value stated in Section~\ref{sec:kr-analysis}.

\section{Summary}
\label{sec:summary}
In this work, we have presented for the first time a dual-phase xenon TPC equipped with SiPMs in the top photosensor array. We operated the detector under stable conditions for more than one year and have demonstrated its performance with data from internal \ce{^{83\text{m}}Kr} and \ce{^{37}Ar} calibration sources.  
Based on the centre of gravity of ionisation signals detected by the SiPM top array, we have reconstructed the $(x,y)$-position with a resolution of $\SI{1.5}{mm}$ within the fiducial radius.
The low energy threshold of the detector and its energy resolution have allowed for an S1 and S2~analysis of the $\SI{2.82}{keV}$ line of \ce{^{37}Ar}. The light and charge yields were compared with NEST~v2.0 predictions, and provide valuable input for the software.
Ongoing studies include an S2-only analysis of the clearly visible $\SI{0.27}{keV}$ line and an attempt to look for the lower energetic M-shell capture at $\SI{17.5}{eV}$. Our study has demonstrated that SiPMs are indeed a promising alternative to PMTs in small detectors.
The scaleability to much larger arrays, such as required by the DARWIN multi-ton liquid xenon detector~\cite{Aalbers:2016jon} is under investigation. A major challenge, apart from reducing the dark count rate,  is the readout and the signal pre-amplification. The main shortcoming of the described amplifier board for 16 channels is its heat dissipation of $\sim \SI{3}{W}$. To that end, we are currently investigating the lowest allowable pre-amplification factor and testing various low-power pre-amplifiers at cryogenic temperatures. 

\begin{acknowledgements}
This work was supported by the European Research Council (ERC) under the European Union's Horizon 2020 research and innovation programme, grant agreement No. 742789 ({\sl Xenoscope}), by the Swiss National Science Foundation under Grant No. 200020-188716, and by the European Unions Horizon 2020 research and innovation programme under the Marie Sklodowska-Curie grant agreements No. 690575 and No. 674896. We thank Reto Maier and the mechanical workshop for the fabrication of the mechanical components for the upgrade including the photosensor holder and the ampule breaking mechanism. We also thank David Wolf and Daniel Florin from the electronics workshop for the design of the SiPM readout board and the signal breakout box and for their continuous help. Furthermore, we thank Daniel Schnarwiler from the glassblowing workshop UZH for producing the source ampules, Alexander V\"{o}gele from the Laboratory of Radiochemistry at PSI for the argon activation and our former group member Alexander Kish for his contributions to the project.

This is a pre-print of an article published in the European Physical Journal C. The final authenticated version is available online at: \href{https://doi.org/10.1140/epjc/s10052-020-8031-6}{https://doi.org/10.1140/epjc/s10052-020-8031-6} 
\end{acknowledgements}

\bibliographystyle{JHEP}
\bibliography{xurich_sipm}

\end{document}